\documentclass[aps, 10pt, prd, twocolumn, superscriptaddress, preprintnumbers, longbibliography, nofootinbib]{revtex4-2}

\usepackage[top=2.8cm, bottom=2.8cm, left=2cm, right=2cm]{geometry}
\usepackage[utf8]{inputenc} 
\usepackage[T1]{fontenc}
\usepackage{amsmath, amssymb, lmodern} 
\usepackage{graphicx}
\usepackage{subfigure}
\usepackage{natbib}
\usepackage{booktabs}
\usepackage{hyperref}
\usepackage{tabularray}
\usepackage{orcidlink}
\usepackage{comment}
\usepackage{cancel}
\hypersetup{colorlinks, citecolor = blue, filecolor = blue, linkcolor = blue, urlcolor = blue}
\usepackage[normalem]{ulem}
\usepackage{xcolor}

\newcommand{\dd}{\text{d}}

\definecolor{darkgreen}{rgb}{0, 0.6, 0}

\begin{document}
\preprint{PI/UAN-2026-738FT}
\title{Phase-Space Constraints on a General Class of Eckart-like Viscous Cosmologies} 

\author{David R. Moreno\orcidlink{0009-0004-8972-0844}}
\email{david2210697@correo.uis.edu.co}
\affiliation{Escuela de Física, Universidad Industrial de Santander, Ciudad Universitaria, Bucaramanga 680002, Colombia}
\author{J. Bayron Orjuela-Quintana\orcidlink{0000-0001-5090-2860}}
\email{john.orjuela@correounivalle.edu.co}
\affiliation{Departamento de Física, Universidad del Valle, Ciudad Universitaria Meléndez, Santiago de Cali 760032, Colombia}
\author{Yeinzon Rodríguez\orcidlink{0000-0002-9830-7128}}
\email{yeinzon@saber.uis.edu.co}
\affiliation{Escuela de Física, Universidad Industrial de Santander, Ciudad Universitaria, Bucaramanga 680002, Colombia}
\affiliation{Centro de Investigaciones en Ciencias Básicas y Aplicadas, Universidad Antonio Nariño, Cra 3 Este \# 47A - 15, Bogotá D.C. 110231, Colombia}

\begin{abstract}
We study the cosmological dynamics of a general class of viscous dark matter models using a dynamical system analysis. In the Eckart formalism, these viscous cosmologies are characterized by a dimensionless bulk viscous pressure $\Omega_\Pi \equiv \Omega_\Pi(\Omega_m)$, where $\Omega_m$ is the matter density parameter. Our analysis demonstrates that a viable early radiation era requires the complete absence of viscosity, a proper matter-dominated epoch demands a small viscous pressure, and the only physically admissible late-time attractor is the cosmological constant-dominated fixed point; consequently, the background expansion must closely track standard $\Lambda$CDM evolution. We validate these general results using the power-law parameterization $\Omega_\Pi = -\hat{\xi}_0 \Omega_m^s$, where $\hat{\xi}_0$ and $s$ are real parameters; for $s \geq 1$ and $\hat{\xi}_0 \ll 1$, the model reproduces the $\Lambda$CDM background, while for $s < 1$ the system encounters a past-time integration boundary where the dark matter density transitions into negative values and its viscous pressure evaluates to nonphysical complex numbers. Since the background expansion essentially duplicates $\Lambda$CDM, the model is unlikely to resolve the $H_0$ tension. However, large differences arise at the linear perturbative level;  indeed, on sub-horizon scales, bulk viscosity causes a scale-dependent dissipative drag on the matter density fluctuations; this increases the Hubble friction, offering a channel to suppress structure growth. 
\end{abstract}

\maketitle

\section{Introduction}
\label{Sec: Intro}

The $\Lambda$CDM model is the standard cosmological paradigm; it provides a simple but accurate description of the Universe's large-scale structure and expansion history~\cite{Peebles:2022bya, Mukhanov:2005sc, Planck:2018vyg}. Despite its profound phenomenological success, this model faces significant theoretical difficulties, most notably the cosmological constant problem~\cite{Weinberg:1988cp, Martin:2012bt}, along with observational tensions and anomalies~\cite{Bull:2015stt, Verde:2019ivm, DiValentino:2020vvd, Perivolaropoulos:2021jda, Abdalla:2022yfr, Watkins:2023rll, DESI:2025zgx}, which continue to motivate the search for alternative mechanisms to explain the current period of accelerated expansion~\cite{Amendola:2015ksp}. Among these alternatives, viscous cosmologies have gained considerable attention~\cite{Velten:2014xca, Hu:2020xus, Villalobos:2025mdk, Bhatt:2019qbq}. These models describe the dark sector with corrections to the energy-momentum tensor caused by microscopic effects that manifest macroscopically through viscosity \cite{PhysRevD.111.083540, Brevik:2017msy}.

From a macroscopic perspective, bulk viscosity arises as the only dissipative mechanism compatible with the fundamental spatial symmetries of a homogeneous and isotropic Friedmann--Lemaître--Robertson--Walker (FLRW) spacetime~\cite{Misner:1967uu, Padmanabhan:1987dg}. Historically, early treatments of cosmic dissipation relied on the first-order Eckart formalism~\cite{Eckart:1940te}; however, this approach is well known to suffer from non-causal signal propagation and gradient instabilities~\cite{Hiscock:1985zz}. These structural pathologies are formally resolved by the Israel--Stewart (IS) theory, a causal second-order thermodynamic framework where the viscous pressure is treated as a fully dynamical variable governed by a characteristic finite relaxation time~\cite{Israel:1976tn, Israel:1979wp, Zakari:1993yhk, Maartens:1995wt, Maartens:1996vi, Wilson:2006gf, Hipolito-Ricaldi:2010wrq, Acquaviva:2014vga}.

In recent decades, an extensive body of literature has explored the possibility of bulk viscosity driving the late-time acceleration of the Universe~\cite{Fabris:2005ts, Colistete:2007xi, Avelino:2008ph, Avelino:2010pb, Hipolito-Ricaldi:2009xbk, Brevik:2011mm, Floerchinger:2014jsa, Sasidharan:2014wqa, Sasidharan:2015ihq, Normann:2016jns, Normann:2016zby, Mohan:2017poq, Atreya:2017pny, Cruz:2017bcv, Cruz:2018psw, Brevik:2020psp, Belinchon:2022vtk}, often reporting competitive fits to observational samples such as Supernovae Ia (SnIa), Cosmic Microwave Background (CMB) anisotropies, and Hubble parameter measurements~\cite{Velten:2011bg, Velten:2012uv,  Brevik:2014eya, Goswami:2016tsu, Mostaghel:2016lcd, Yang:2019qza, Cruz:2019uya, Hernandez-Almada:2021osl, Gomez:2022qcu, Cruz:2023dzn, Palma:2024qrw, Palma:2025qge, Adil:2026ift}, or shedding light on observational tensions~\cite{Anand:2017wsj, Elizalde:2020mfs, Normann:2021bjy}. Remarkably, however, the vast majority of these works focus almost exclusively on the late-time matter-to-dark-energy transition. In doing so, they frequently truncate the expansion history by entirely omitting the radiation component from their dynamical equations or by initiating numerical integrations well within the matter-dominated epoch. This widespread neglect overlooks an essential condition for any feasible cosmological model: it must not only produce acceleration today, but it must also self-consistently preserve a stable, primordial radiation-dominated era to protect early-universe observables, such as Big Bang Nucleosynthesis (BBN), before transitioning into an adequately prolonged matter epoch~\cite{Gagnon:2011id}. 

In this paper, we perform a systematic analysis of viscous cosmologies, within the Eckart framework, characterized by a dimensionless viscous pressure  parameterized in general as $\Omega_\Pi = \Omega_\Pi(\Omega_m)$, where $\Omega_m$ is the matter density parameter. By implementing a dynamical system approach, we establish phase-space boundaries on this general model including the radiation fluid. We demonstrate that the requirement of a viable radiation-dominated source node severely restricts the background dynamics, forcing the expansion history to closely duplicate that of standard $\Lambda$CDM.

Crucially, this background degeneracy does not imply that the model is phenomenologically inert. In agreement with previous literature exploring viscous dark matter models~\cite{Velten:2012uv, Velten:2013pra, Velten:2014xca}, evaluating the system at the linear perturbative level under the sub-horizon and quasi-static approximations reveals that the bulk viscosity introduces a scale-dependent dissipative drag. This well-known mechanism modulates the Hubble friction, providing an active channel to suppress structure growth that has been proposed as a remedy for the $\sigma_8$ (or $S_8$) tension~\cite{Anand:2017wsj, Ashoorioon:2023jwf}. In this context, the primary contribution of our study is not the novelty of this perturbative suppression, but rather the discovery that its viability is strictly bounded by cosmic chronology. By demonstrating that large regions of the parameter space may trigger unphysical past-time singularities, we establish a restriction on the feasibility of the models at the background level, recontextualizing the conditions under which the class of viscous cosmologies considered in this paper can safely operate as a subtle extension of the standard $\Lambda$CDM model.

The paper is organized ain the following way. In Sec.~\ref{Sec: Viscous DM}, we review the fundamental contributions to the energy-momentum tensor introduced by the bulk viscosity. In Sec.~\ref{Sec: DS Analysis}, we introduce the dynamical systems setup and derive phase-space constraints dictated by the early-universe fixed points. Section~\ref{Sec: Case Study} is devoted to a detailed case study of the power-law parameterization, including its local stability, numerical integration, and the identification of past-time mathematical boundaries. We also develop the linear perturbation framework and discuss its implications for large-scale structures. Finally, in Sec.~\ref{Sec: Conclusions}, we summarize our conclusions and outline future perspectives.

\section{Viscous Dark Matter}
\label{Sec: Viscous DM}

In the $\Lambda$CDM model, dark matter is modeled as a perfect, pressureless fluid. However, self-interactions or the continuous decay of dark matter particles into lighter species can give rise to dissipative effects at the macroscopic level~\cite{Spergel:1999mh, Tulin:2017ara}. Under the assumption of a homogeneous and isotropic universe, shear viscosity vanishes, leaving bulk viscosity as the sole dissipative mechanism. In the following, we employ the formalism developed in Ref.~\cite{Zimdahl:1996ka}. 

The energy-momentum tensor for a fluid with bulk viscosity is given by:
\begin{equation}
    T_{\mu\nu} = \rho \, u_\mu u_\nu + (p + \Pi) h_{\mu\nu},
\end{equation}
where $\rho$ is the energy density, $p$ is the equilibrium thermodynamic pressure, $u_\mu$ is the fluid four-velocity, and $\Pi$ is the bulk viscous pressure of the fluid. The projection tensor onto the orthogonal subspace to the fluid flow is defined as $h_{\mu\nu} \equiv g_{\mu\nu} + u_\mu u_\nu$, where $g_{\mu \nu}$ is the metric tensor.

The causal theory to describe the evolution of $\Pi$ relies on the relativistic thermodynamics of Israel and Stewart \cite{Israel:1976tn,Israel:1979wp}. Up to second order in the dissipative fluxes, the particle number current $N^\alpha$ and the entropy current $S^\alpha$ are defined as:
\begin{equation}
\label{Eq: Entropy Current}
    N^\alpha \equiv n \, u^\alpha, \quad S^\alpha \equiv s \, N^\alpha - \frac{\tau_\text{r} \, \Pi^2}{2 \, \xi \, T} u^\alpha,
\end{equation}
where $n$ is the particle number density, $s$ is the entropy per particle, $T$ is the fluid temperature, $\xi$ is the bulk viscosity coefficient, and $\tau_\text{r}$ is the relaxation time that characterizes the time delay for the system to reach its local thermodynamic equilibrium state.

The covariant conservation of the particle number ($\nabla_\mu N^\mu = 0$) and the energy-momentum tensor ($\nabla_\mu T^{\mu\nu} = 0$) yield the standard continuity equations:
\begin{equation}
\label{Eq: Continuity Eqs}
    \dot{n} = -\Theta \,n, \quad \dot{\rho} = - \Theta (\rho + p + \Pi),
\end{equation}
where $\Theta \equiv \nabla_\alpha u^{\alpha}$ is the fluid expansion scalar, and the overdot denotes the absolute cosmic-time derivative $\dot{X} \equiv u^\alpha \partial_\alpha X$. To those equations, we also add the Gibbs--Duheim equation for the evolution of the entropy per particle:
\begin{equation}
\label{Eq: Gibbs-Duheim Eq}
    n \, T \, \dot{s} = - \Theta \, T.
\end{equation}

Taking the divergence of the entropy current in Eq.~\eqref{Eq: Entropy Current} and using the continuity equations in Eq.~\eqref{Eq: Continuity Eqs} and \eqref{Eq: Gibbs-Duheim Eq}, the entropy production rate reads:
\begin{equation}
    \nabla_\alpha S^\alpha = - \frac{\Pi}{T} \left[ \Theta + \frac{\tau_\mathrm{r}}{\xi} \dot{\Pi} + \frac{1}{2}  \Pi \, T \, \nabla_\alpha\left( \frac{\tau_\mathrm{r}}{T \, \xi} u^\alpha \right)\right].
\end{equation}
The second law of thermodynamics strictly requires the entropy production to be non-negative, $\nabla_\alpha S^\alpha \geq 0$. To satisfy this inequality unconditionally, we invoke the standard linear phenomenological law of non-equilibrium thermodynamics, which dictates that the thermodynamic force---represented by the bracketed expression---is directly proportional to the dissipative flux ($-\Pi$) through a positive transport coefficient. Following the standard IS procedure, we obtain the evolution equation for the bulk viscous pressure:
\begin{equation}
\label{Eq: Full Pi eq}
    \Pi + \tau_\mathrm{r} \, \dot{\Pi} = - \xi \Theta - \frac{\tau_\mathrm{r}}{2} \Pi \left[ \Theta + \frac{\dot{\tau_\mathrm{r}}}{\tau_\mathrm{r}} - \frac{\dot{\xi}}{\xi} - \frac{\dot{T}}{T} \right].
\end{equation}
In this formalism, the entropy production takes the quadratic form:
\begin{equation}
    \nabla_\alpha S^\alpha = \frac{\Pi^2}{T \, \xi} \geq 0 \quad \implies \quad \xi > 0.
\end{equation}
This is a fundamental thermodynamic constraint: the bulk viscosity coefficient $\xi$ must be strictly positive to prevent a violation of the second law.

This full causal formalism possesses two relevant limits. The truncated version of the IS theory is given by the case when the bracket terms on the right-hand side of the Eq.~\eqref{Eq: Full Pi eq} are negligible in comparison with the viscosity term:
\begin{equation}
\label{eq: Trunc Pi eq}
    \Pi + \tau_\mathrm{r} \,\dot{\Pi} = -\xi\Theta.
\end{equation}
This approximation is valid when the system is near the equilibrium state and, in some exceptional cases, is identical to the full version~\cite{Zimdahl:1996ka, Maartens:1996vi}. This truncated version has some advantages: it avoids complicated nonlinear terms, predicts a finite speed for the viscous pulses as in the complete framework, and could be compatible with astronomical observations~\cite{Mathew:2020jjx}.

In the limit where the relaxation time is vanishingly small ($\tau_\text{r} \to 0$), the theory reduces to the non-causal Eckart framework, yielding the algebraic relation:
\begin{equation}
\label{Eq: Eckart Pi}
    \Pi = -\xi \Theta.
\end{equation}

In the Eckart framework, the bulk viscous pressure $\Pi$ takes on negative values in an expanding universe ($\Theta > 0$) regardless of the specific parameterization employed for $\xi$. In the truncated IS theory, $\Pi$ also tends to negative values in some specific cases, as we show in Appendix~\ref{App: Truncated IsrSte}. 

As we will demonstrate through a phase-space analysis, this negative pressure—while phenomenologically appealing for accelerating the late-time expansion—forces the Eckart-like viscous dynamics to closely track the background history of the standard cosmological paradigm.

\section{Dynamical System Analysis}
\label{Sec: DS Analysis}

To evaluate the viability of viscous cosmologies, we recast the equations of motion as an autonomous dynamical system, whose fixed points and stability determine their global asymptotic behaviour~\cite{Bahamonde:2017ize}.

\subsection{Autonomous Set}

We assume an isotropic and homogeneous universe described by an FLRW metric containing radiation and viscous dark matter with densities $\rho_r$ and $\rho_m$, respectively, and a cosmological constant $\Lambda$. The dynamics is encoded in the Friedmann equations:
\begin{align}
    H^2 &= \frac{8\pi G_\mathrm{N}}{3}(\rho_r + \rho_m) + \frac{\Lambda}{3}, \\
    2\dot{H} + 3H^2 &= -8\pi G_\mathrm{N} \left(\frac{1}{3}\rho_r + \Pi \right)+ \Lambda,
\end{align}
where $H$ is the Hubble rate and $G_\mathrm{N}$ is the Newton's constant. They are complemented with the continuity equations:
\begin{align}
    \dot{\rho}_r + 4H\rho_r=0, \\
    \dot{\rho}_m + 3H(\rho_m + \Pi) = 0.
\end{align}

We define the dimensionless density parameters as:
\begin{equation}
    \Omega_r \equiv \frac{8\pi G}{3 H^2} \rho_r, \quad \Omega_m \equiv \frac{8\pi G}{3 H^2} \rho_m, \quad \Omega_\Lambda \equiv \frac{\Lambda}{3 H^2},
\end{equation}
such that the Friedmann equation yields the algebraic constraint
\begin{equation}
    \Omega_r + \Omega_m + \Omega_\Lambda = 1,
\end{equation}
which bounds the physical phase space. 

Choosing the number of $e$-folds, $N \equiv \ln a$, as the independent time variable, and employing the continuity equations along with the second Friedmann equation, the cosmological dynamics can be expressed as an autonomous system:
\begin{align}
    \frac{\dd \Omega_r}{\dd N} &= 2 \Omega_r (q - 1), \label{eq: Radiation evolution}\\
    \frac{\dd \Omega_m}{\dd N} &= (2q - 1) \Omega_m - 3 \Omega_\Pi, \label{eq: Matter evolution}
\end{align}
where $\Omega_\Lambda$ has been eliminated via the Friedmann constraint, and we have defined the dimensionless bulk viscous pressure as:
\begin{equation}
\label{Eq: Omegapi}
    \Omega_\Pi \equiv \frac{8\pi G}{3 H^2} \Pi.
\end{equation}
The deceleration parameter $q \equiv - a \ddot{a}/\dot{a}^2$ is expressed in terms of the dimensionless variables as:
\begin{equation}
    q = -1 + 2\Omega_r + \frac{3}{2}(\Omega_m + \Omega_\Pi).
\end{equation}

In the following, we impose the following physically plausible conditions on $\Omega_\Pi$: we express it solely as a function of the matter density parameter, i.e., $\Omega_\Pi = \Omega_\Pi(\Omega_m)$, and $\Omega_\Pi = 0$ when $\Omega_m = 0$, which means that there is no bulk viscous pressure when there is no viscous matter. These assumptions reduce the dimension of the phase space to two degrees of freedom.

\subsection{Fixed Points}

The critical points of the autonomous system are found by setting:
\begin{equation}
\frac{\dd \Omega_r}{\dd N }= 0, \quad \frac{\dd \Omega_m}{\dd N }= 0.
\end{equation}
The former condition requires:
\begin{equation}
    2 \Omega_r (q - 1) = 0.
\end{equation}
This result yield to two mutually exclusive branches for the radiation sector:
\begin{equation}
    \text{(B1)} \quad \Omega_r = 0, \qquad \text{(B2)} \quad q = 1.
\end{equation}
These branches set up the routes for classifying the fixed points of the entire system.

\subsubsection{\textbf{First Branch (B1)}}

The branch $\Omega_r = 0$ (B1) corresponds to the late-time universe, when the radiation fluid has decoupled and diluted. Under this condition, the deceleration parameter becomes:
\begin{equation}
    q = -1 + \frac{3}{2}(\Omega_m + \Omega_\Pi).
\end{equation}
Substituting this relation into the matter evolution equation~\eqref{eq: Matter evolution} yields the reduced system:
\begin{equation}
    \left.\frac{\dd\Omega_m}{\dd N}\right|_{\text{(B1)}} = - 3 (1 - \Omega_m) (\Omega_m + \Omega_\Pi).
\end{equation}
Setting this derivative to zero yields three different fixed points for the late-time cosmic dynamics:
\begin{equation*}
    \text{(M)} \quad \Omega_m = 1, \qquad \text{(VM)} \quad \Omega_m = - \Omega_\Pi,
\end{equation*}
\begin{equation}
    (\Lambda) \quad \Omega_m = 0.
\end{equation}

These solutions correspond to different physical regimes:
\begin{enumerate}
    \item \textbf{Matter-Dominated Point (M)}: The solution with $\Omega_m = 1$ corresponds to the standard matter-dominated epoch ($\Omega_\Lambda = 0$). At this point, the effective equation of state\footnote{The effective equation of state is related to the deceleration parameter by $w_\text{eff} = (2q -1)/3$.} is $w_\text{eff} = \Omega_\Pi$. The usual evolution of pressureless dust is preserved whenever $|\Omega_\Pi| \ll 1$. 
    
    \item \textbf{Viscous Matter-Dominated Point (VM)}: This point is defined by $\Omega_m = - \Omega_\Pi$. For physical consistency, $\Omega_m > 0$, the bulk viscous pressure must be $\Omega_\Pi < 0$. This solution yields a deceleration parameter $q = -1$ (or $w_{\text{eff}} = -1$), corresponding to a de Sitter expansion phase. Thus, the (VM) configuration acts as a dark energy surrogate driven entirely by non-equilibrium dissipative mechanisms.
    
    \item \textbf{de Sitter Point ($\Lambda$):} Defined by $\Omega_m = 0$, which implies $\Omega_\Lambda = 1$ via the Friedmann constraint. This point represents the standard late-time attractor where the cosmological constant completely takes over the cosmic energy budget.
\end{enumerate}

\subsubsection{\textbf{Second Branch (B2)}}

We now examine the branch (B2) defined by the condition $q = 1$ that results in a background dynamics driven by an effective equation of state $w_{\text{eff}} = 1/3$. Therefore, this branch represents the phase-space region associated with the early radiation era.

Substituting $q = 1$ into Eq.~\eqref{eq: Matter evolution} and setting $\dd \Omega_m/\dd N = 0$ gives the fixed point:
\begin{equation}
    \Omega_m|_{\text{(B2)}} = 3 \Omega_\Pi.
\end{equation}
This condition imposes severe restrictions on the dark matter sector at early times:
\begin{enumerate}
    \item \textbf{Radiation-like scaling}: The constraint relates the bulk viscous pressure to the matter energy density through the equation $\Pi = \frac{1}{3} \rho_m$, similar to a radiation-like fluid.
    
    \item \textbf{Positivity of viscous pressure}: A positive dark matter energy density ($\Omega_m \geq 0$) requires that the dimensionless bulk viscous pressure must satisfy $\Omega_\Pi \geq 0$.
\end{enumerate}

This result introduces tension with the thermodynamics established in Sec.~\ref{Sec: Viscous DM}. Since the second law of thermodynamics requires $\xi > 0$, the Eckart transport in Eq.~\eqref{Eq: Eckart Pi} drives the bulk viscous pressure toward negative equilibrium values in an expanding universe, and thus $\Omega_\Pi \leq 0$. Furthermore, a standard radiation era is characterized by $\Omega_r \approx 1$. However, the Friedmann constraint together with the equilibrium condition implies $\Omega_r = 1 - 3\Omega_\Pi$, and thus the viscous dark matter contribution must be negligible but positive to avoid $\Omega_r$  overshooting unity. 

As a result, the phase-space analysis reveals that a physically viable radiation era cannot be sustained alongside a viscous dark matter component under consistent thermodynamics conditions. Hence, the fixed point
\begin{equation}
    \text{(R1)} \quad \Omega_r = 1-3\Omega_\Pi, \quad \Omega_m = 3 \Omega_\Pi,
    \end{equation}
becomes stable and thermodynamically consistent only when:
\begin{equation}
    \text{(R)} \quad \Omega_r = 1, \quad \Omega_m = 0, \quad \Omega_\Pi = 0,
    \end{equation}
which corresponds to the standard radiation-dominated point of the $\Lambda$CDM model.

\subsection{Stability Analysis}

A successful cosmological trajectory must reproduce the following expansion history: it should begin in a primordial radiation-dominated source (repeller point), pass through an intermediate matter-dominated stage (saddle point), and then move towards a late-time accelerated era (attractor point). We characterize this sequence, by performing a linear stability analysis around each fixed point.

We define the dynamical system in terms of the vector components $f_m \equiv \dd\Omega_m/\dd N$ and $f_r \equiv \dd\Omega_r/\dd N$. The stability of the system in the neighborhood of any fixed point is determined by the eigenvalues of the $2 \times 2$ Jacobian matrix defined by $J_{ij} \equiv \partial f_i/\partial \Omega_j$, where:
\begin{align}
    J_{11} &= (2q - 1) + 2\Omega_m \left(\frac{\partial q}{\partial \Omega_m}\right) - 3 \left(\frac{\partial \Omega_\Pi}{\partial \Omega_m}\right), \\
    J_{12} &= 2 \Omega_m \left(\frac{\partial q}{\partial \Omega_r}\right), \\
    J_{21} &= 2 \Omega_r \left(\frac{\partial q}{\partial \Omega_m}\right), \\
    J_{22} &= 2(q - 1) + 2\Omega_r \left(\frac{\partial q}{\partial \Omega_r}\right).
\end{align}
The derivatives of the deceleration parameter are given by:
\begin{equation}
    \frac{\partial q}{\partial \Omega_m} = \frac{3}{2} \left(1 + \frac{\dd \Omega_\Pi}{\dd \Omega_m} \right), \quad \frac{\partial q}{\partial \Omega_r} = 2.
\end{equation}

A critical point acts as a source (repeller or unstable node) when the real parts of both eigenvalues are strictly positive. It constitutes an unstable saddle point when the real parts of the eigenvalues have opposite signs. Finally, the point behaves as an attractor (sink or stable node) when the real parts of both eigenvalues are strictly negative.

\subsubsection{\textbf{The Repeller Condition for (R1)}}

\begin{figure}[t]
    \centering
    \includegraphics[width=0.9\linewidth]{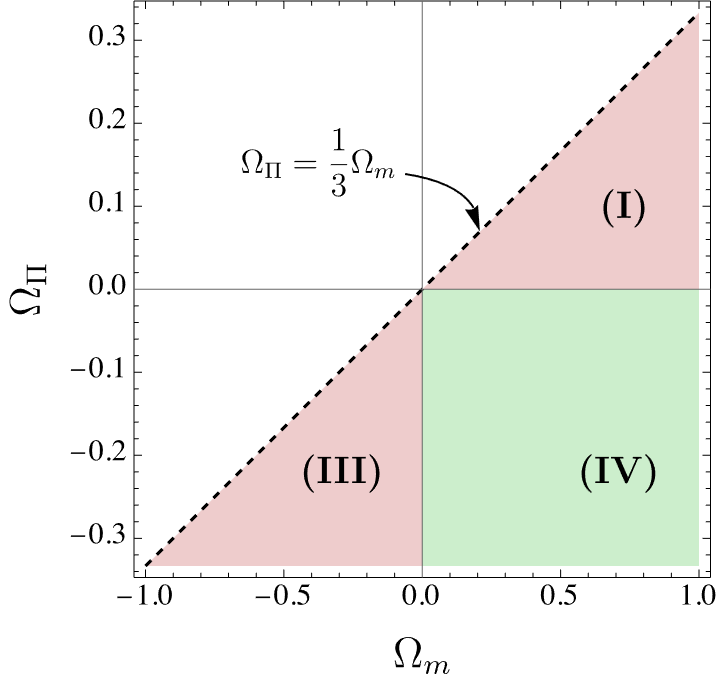}
    \caption{Illustration of the $\Omega_m$--$\Omega_\Pi$ phase plane. The dashed black line represents the equilibrium condition $\Omega_\Pi = \frac{1}{3}\Omega_m$, which is required for the existence of the radiation-dominated critical configuration (R1). Region (IV) (highlighted in green) corresponds to the thermodynamically and physically viable parameter space, described by $\Omega_m > 0$ and $\Omega_\Pi \leq 0$. Regions (I) and (III), shaded in red, correspond to unphysical domains associated with thermodynamic violations ($\nabla_\alpha S^\alpha < 0$) or negative matter energy density, respectively. The absence of an equilibrium state in Region (IV) demonstrates the phase-space restrictions for sustaining a viscous radiation era under standard thermodynamic conditions.}
    \label{Fig: Repeller R2}
\end{figure}

Additional to the physical unviability of (R1), in the following we show that cosmic trajectories cannot consistently start from it. 

For a two-dimensional system, the eigenvalues $\lambda_1$ and $\lambda_2$ of the Jacobian matrix can be used to compute the determinant and its trace:
\begin{equation}
    \det(J) = \lambda_1 \lambda_2, \quad \text{tr}(J) = \lambda_1 + \lambda_2.
\end{equation}
For the fixed point (R1) to serve as the origin of physical trajectories, the Jacobian matrix associated with it must obey $\det(J) > 0$ and $\text{tr}(J) > 0$. Evaluating the determinant at (R1), where $q = 1$, we find:
\begin{equation}
    \det(J)|_\text{(R1)} = 4 \Omega_r \left(1 - 3 \frac{\dd \Omega_\Pi}{\dd \Omega_m} \right).
\end{equation} 
Given that $\Omega_r > 0$, (R1) acts as a repeller provided that:
\begin{equation}
    \left. \frac{\dd \Omega_\Pi}{\dd \Omega_m} \right|_\text{(R1)} < \frac{1}{3}.
\end{equation}
However, the equilibrium configuration for (R1) requires that the system be situated exactly on the locus $\Omega_\Pi = \frac{1}{3}\Omega_m$. As shown in Fig.~\ref{Fig: Repeller R2}, the linear relation $\Omega_\Pi = \frac{1}{3}\Omega_m$ applies only to Region (I) ($\Omega_m > 0, \Omega_\Pi > 0$) and Region (III) ($\Omega_m < 0, \Omega_\Pi < 0$). Conversely, the physically allowed domain (Region (IV)) is strictly given by $\Omega_m > 0$ (energy positivity) and $\Omega_\Pi \leq 0$ (compliance with the second law in an expanding background).

Since the equilibrium curve of (R1) does not cross Region (IV), it is impossible for a trajectory to start from this viscous state and simultaneously satisfy standard thermodynamic requirements. Consequently, any trajectory with a standard radiation-dominated background must start from the non-viscous limit, $\Omega_\Pi = 0$.

\subsubsection{\textbf{Stability of Global Fixed Points}}

Under the condition where the dimensionless viscous pressure vanishes alongside the matter density parameter ($\Omega_\Pi \to 0$ as $\Omega_m \to 0$), the fixed points of the system are characterized as follows:
\begin{itemize}
    \item \textbf{Radiation Point (R):} It is defined by $\Omega_r = 1$. The associated eigenvalues are $\lambda_1 = 4$ and $\lambda_2 = 1$, which means that this point is the source of the system.
    
    \item \textbf{Matter Point (M):} This point is characterized by $\Omega_m = 1$, with eigenvalues $\lambda_1 = -1 + 3\Omega_\Pi$ and $\lambda_2 = 3(1+\Omega_\Pi)$. When $|\Omega_\Pi| < 1/3$, (M) is a saddle point and let the system transit from an early radiation era to a late-time cosmological evolution.
    
    \item \textbf{Viscous Matter Point (VM):} Defined by the equilibrium condition $\Omega_m = -\Omega_\Pi$, with eigenvalues $\lambda_1 = -4$ and $\lambda_2 = 0$. The presence of a zero eigenvalue indicates that this configuration is non hyperbolic. Consequently, the linear theory is inconclusive and other techniques are required to assess the stability of the fixed point~\cite{Bahamonde:2017ize}.
    
    \item \textbf{de Sitter Point ($\Lambda$):} With the condition $\Omega_\Lambda = 1$, and eigenvalues $\lambda_1 = -4$ and $\lambda_2 = -3$, this point represents the system's late-time attractor.
\end{itemize}

The stability analysis shows that while the late-time accelerated expansion can be driven by either the dissipative viscous mechanism (VM) or the cosmological constant ($\Lambda$), the early radiation era is strongly constrained in receiving sustained viscous contributions. At the $q = 1$ threshold, any non-zero bulk viscosity either violates the second law or forces the dark matter energy density into unacceptable negative values. Moreover, $\Omega_\Pi$ is tightly constrained throughout the matter-dominated epoch to preserve the necessary saddle-point transition, and the final stage of the system is determined by the stable de Sitter node. As a result, throughout cosmic history, the background evolution remains close to the $\Lambda$CDM trajectory.

\subsection{Structural Generality and Background Indistinguishability}
\label{Sec: Generality}

In the Eckart formalism, bulk viscous pressure is characterized by the bulk viscosity coefficient $\xi$ and the expansion scalar. Therefore, we believe that our general parameterization $\Omega_\Pi \equiv \Omega_\Pi(\Omega_m)$, with $\Omega_\Pi \to 0$ when $\Omega_m \to 0$ represents a physically well-motivated and broadly general assumption.

The limitations identified in the phase-space analysis arise from the theoretical framework rather than from a specific parameterization for $\Pi$. These restrictions stem from four fundamental conditions: the negativity of $\Pi$ dictated by Eckart theory, the fluid continuity equations, the Friedmann algebraic constraint, and the thermodynamic requirement of non-negative entropy production. Since those features do not depend on the specific form of $\Pi$ or $\xi$, our findings possess a high degree of generality and point out that, within the conditions presented, viscous dark matter models evolve almost indistinguishably from the standard $\Lambda$CDM background.

As shown in Appendix~\ref{App: Truncated IsrSte}, our assumption $\Omega_\Pi \equiv \Omega_\Pi(\Omega_m)$ implies that $\Pi$ is driven toward negative values in the truncated IS formalism as well. However, further analysis is required to demonstrate, in full generality, that this causal formalism suffers from pathologies similar to those of the Eckart formalism.

\section{Case Study: A Specific parameterization}
\label{Sec: Case Study}

To explicitly illustrate the structural constraints obtained from the general phase-space analysis, we apply the dynamical system to a concrete, widely studied phenomenological model.

A standard ansatz in the literature assumes that the bulk viscosity coefficient follows a power-law relation scaled by the Hubble parameter~\cite{Gomez:2022qcu}:
\begin{equation}
\label{Eq: xi Power-Law}
    \xi = \frac{\hat{\xi}_0}{8\pi G} H \Omega_m^s,
\end{equation}
where $\hat{\xi}_0$ and $s$ are constant parameters.
Using the Eckart closure relation ($\Pi = -3H\xi$), the dimensionless bulk viscous pressure takes the form:
\begin{equation}
    \Omega_\Pi = - \hat{\xi}_0 \Omega_m^s.
\end{equation}

To ensure that this phenomenological model is physically and mathematically well behaved, we impose the following bounds on these parameters:
\begin{itemize}
    \item \textbf{Thermodynamic Consistency:} As dictated by the second law, $\xi > 0$ requires $\hat{\xi}_0 > 0$, which consequently implies $\Omega_\Pi \leq 0$ across the physical phase space.
    
    \item \textbf{Asymptotic Regularity ($s > 0$):} To avoid divergences in the viscous pressure as the matter density parameter tends to zero, it is necessary that $s > 0$. This ensures that $\Omega_\Pi \to 0$ as $\Omega_m \to 0$, which means that there is no bulk viscous pressure when viscous matter vanishes.
\end{itemize}
With this parameterization, the general fixed-point constraints and stability conditions derived in Sec.~\ref{Sec: DS Analysis} can be analytically evaluated.

\subsection{Fixed Points}

Applying the power-law parameterization to our general equilibrium conditions, we can analytically assess the restricted phase space and characterize the accessible cosmic epochs.

Let us first examine the radiation-like point (R1). As established in Sec.~\ref{Sec: DS Analysis}, this configuration is governed by the strict algebraic constraint $\Omega_m = 3 \Omega_\Pi$. Substituting the viscous ansatz into this relation yields:
\begin{equation}
    \Omega_m + 3\hat{\xi}_0 \Omega_m^s = 0.
\end{equation}
Because a physical matter density requires $\Omega_m > 0$, and thermodynamic consistency enforces $\hat{\xi}_0 > 0$, the left-hand side of this expression is strictly positive. Consequently, the equation admits no real, positive roots.\footnote{Setting aside the trivial case $\Omega_m = 0$, solving analytically for the matter density yields the formal root $\Omega_m = (-3\hat{\xi}_0)^{1/(1-s)}$. For $\hat{\xi}_0 > 0$, this expression evaluates to either a complex number or a strictly negative value, depending on the domain of $s$. This confirms that the algebraic manipulation introduces extraneous roots that are not present in the physical domain $\Omega_m > 0$.} This explicit analytical contradiction confirms that the (R1) fixed point resides entirely outside the physically accessible phase space.

With the (R1) configuration excluded, we evaluate the remaining critical points:
\begin{itemize}
    \item \textbf{Radiation (R) and de Sitter ($\Lambda$) Points:} The condition $s > 0$ ensures that $\Omega_\Pi \to 0$ as $\Omega_m \to 0$. Therefore, the pure radiation era ($\Omega_r=1$, $\Omega_m=0$) and the cosmological constant-dominated era ($\Omega_r=0$, $\Omega_m=0$) remain dynamically equivalent to their standard $\Lambda$CDM counterparts.
    
    \item \textbf{Matter-Dominated Point (M):} Located at $\Omega_m = 1$, at this equilibrium point we have $\Omega_\Pi = -\hat{\xi}_0$ and $w_{\text{eff}} = -\hat{\xi}_0$. Therefore, this point approximates the standard pressureless dust regime when the bulk viscous parameter is small ($\hat{\xi}_0 \ll 1$).
    
    \item \textbf{Viscous Matter Point (VM):} The equilibrium condition $\Omega_m = -\Omega_\Pi$ reduces to $\Omega_m = \hat{\xi}_0 \Omega_m^s$. At this fixed point:
    \begin{equation}
        \Omega_m = \hat{\xi}_0^{1/(1-s)}, \quad w_{\text{eff}} = -1.
    \end{equation}
    This indicates that the (VM) configuration can act as a fully dissipative-driven de Sitter phase, mimicking the late-time accelerated expansion.
\end{itemize}

\subsection{Stability Analysis} \label{sa}

Evaluating the Jacobian matrix for the power-law parameterization shows that the system's stability is highly sensitive to the scaling exponent $s$. Substituting the ansatz $\Omega_\Pi = -\hat{\xi}_0 \Omega_m^s$ into the general eigenvalue expressions, we characterize the local stability properties of the critical points:

\begin{itemize}
    \item \textbf{Radiation Point (R):} At this point the eigenvalues are given by:
    \begin{equation}
        \lambda_1 = 4, \quad \lambda_2 = 1 + 3s \hat{\xi}_0 \Omega_m^{s - 1}.
    \end{equation}
    Near this point $(\Omega_m \to 0)$, although the positive value of $\lambda_1$ confirms the repeller character of this configuration, the behaviour of $\lambda_2$ depends strongly on the choice of $s$. In the range $0 < s < 1$, the second eigenvalue diverges ($\lambda_2 \to +\infty$) as the trajectory gets close to this point. This divergence shows that the linear approximation breaks down. Thus, it is necessary $s \geq 1$.

    \item \textbf{Matter-Dominated Point (M):} At this equilibrium configuration ($\Omega_m = 1$), the local eigenvalues reduce to:
    \begin{equation}
        \lambda_1 = -1 - 3\hat{\xi}_0, \quad \lambda_2 = 3(1 - \hat{\xi}_0).
    \end{equation}
    This phase can act as a transition epoch, whenever it conserves a saddle-point structure ($\lambda_1 < 0$ and $\lambda_2 > 0$). This dynamical requirement is satisfied as long as $\hat{\xi}_0 < 1$. 

    \item \textbf{Viscous Matter Point (VM):} The determinant of the Jacobian matrix at this point evaluates to:
    \begin{equation}
        \det(J) = 12(1 - s)(1 - \Omega_m).
    \end{equation}
    For (VM) to serve as a stable late-time attractor, the determinant must be strictly positive, which implies $s < 1$ throughout the physical domain $\Omega_m < 1$. On the other hand, if $s > 1$, then the equilibrium value $\Omega_m = \hat{\xi}_0^{1/(1-s)}$ becomes greater than the unity when $\hat{\xi}_0 < 1$. Consequently, under these parametric conditions, the (VM) attractor lies outside the region of the physically allowable phase space ($\Omega_m \in [0, 1]$), and thus becomes physically inaccessible to cosmic trajectories.

    \item \textbf{de Sitter Point ($\Lambda$):} The eigenvalues at this point are:
    \begin{equation}
        \lambda_1 = -4, \quad \lambda_2 = -3 + 3s \hat{\xi}_0 \Omega_m^{s-1}.
    \end{equation}
    Mirroring the behaviour observed at the radiation point, the second eigenvalue diverges ($\lambda_2 \to +\infty$) if $0 < s < 1$ as $\Omega_m \to 0$, thereby disrupting the linear stability of the de Sitter attractor. It follows that in order for ($\Lambda$) to behave as a well-defined, stable late-time sink within the linear approximation, the model requires $s > 1$.
\end{itemize}

\subsection{Parametric Tensions and Phase-Space Restrictions}

The stability analysis shows that there are strict boundaries in the phase space of the power-law formulation. To maintain well-behaved linear stability properties at the past and future boundaries—such that neither the radiation source (R) nor the de Sitter attractor ($\Lambda$) exhibits divergent eigenvalues—requires $s > 1$. However, enforcing this condition restricts the viscous matter point (VM) from residing within the physically allowed region of the phase space. 

Because of this, the present case study illustrates that the framework cannot, using the (VM) configuration, simultaneously achieves late-time acceleration through bulk viscous pressure and maintain the dynamical stability. Instead, the viscous model closely follows the cosmological evolution of the standard $\Lambda$CDM, placing strong constraints on any manifestation of viscous dark matter at the background level. 

\subsection{Numerical Integration}

\begin{figure*}[t]
    \centering
    {\includegraphics[width=0.35\linewidth]{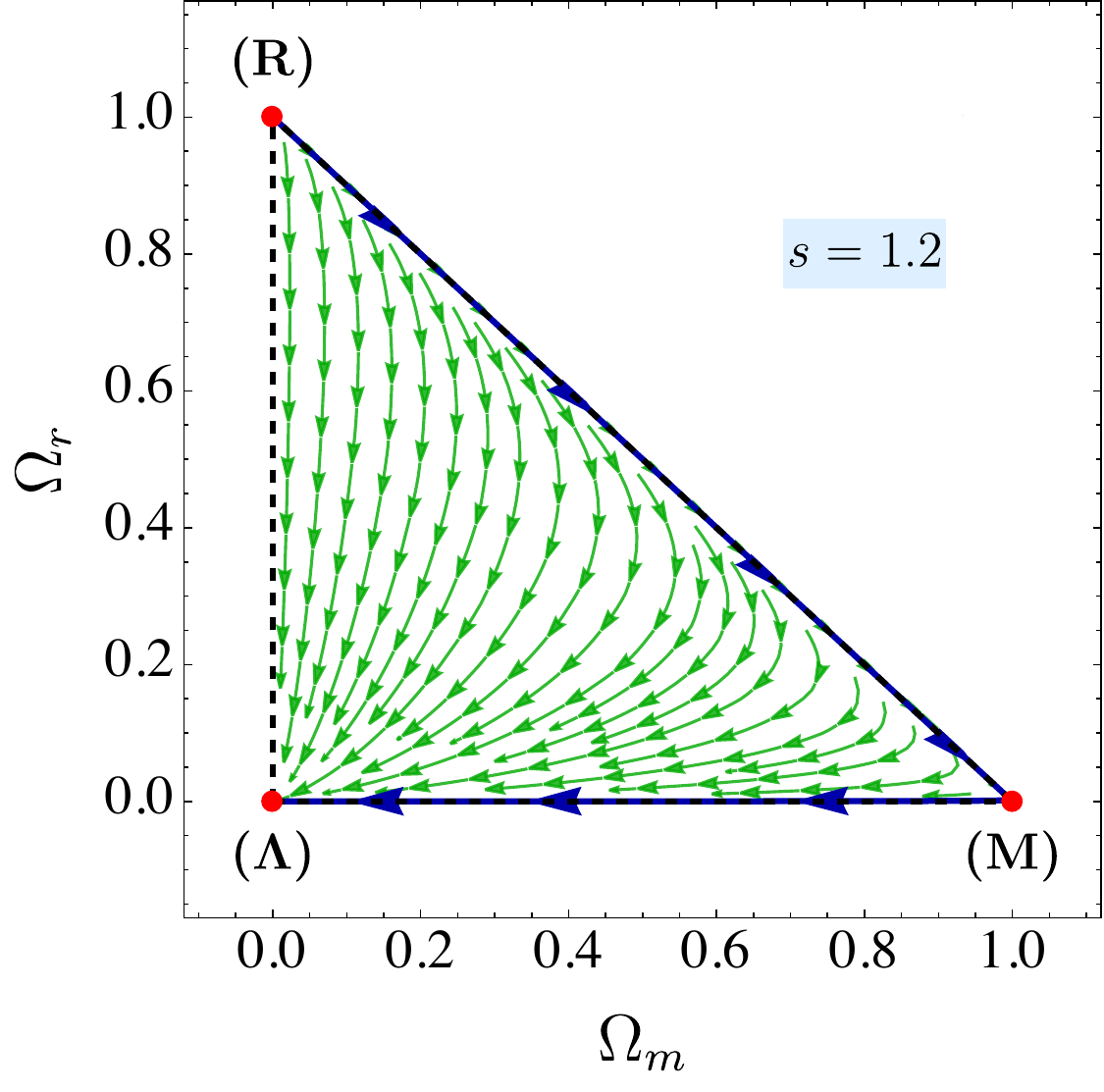} \hspace{2cm}
    \includegraphics[width=0.35\linewidth]{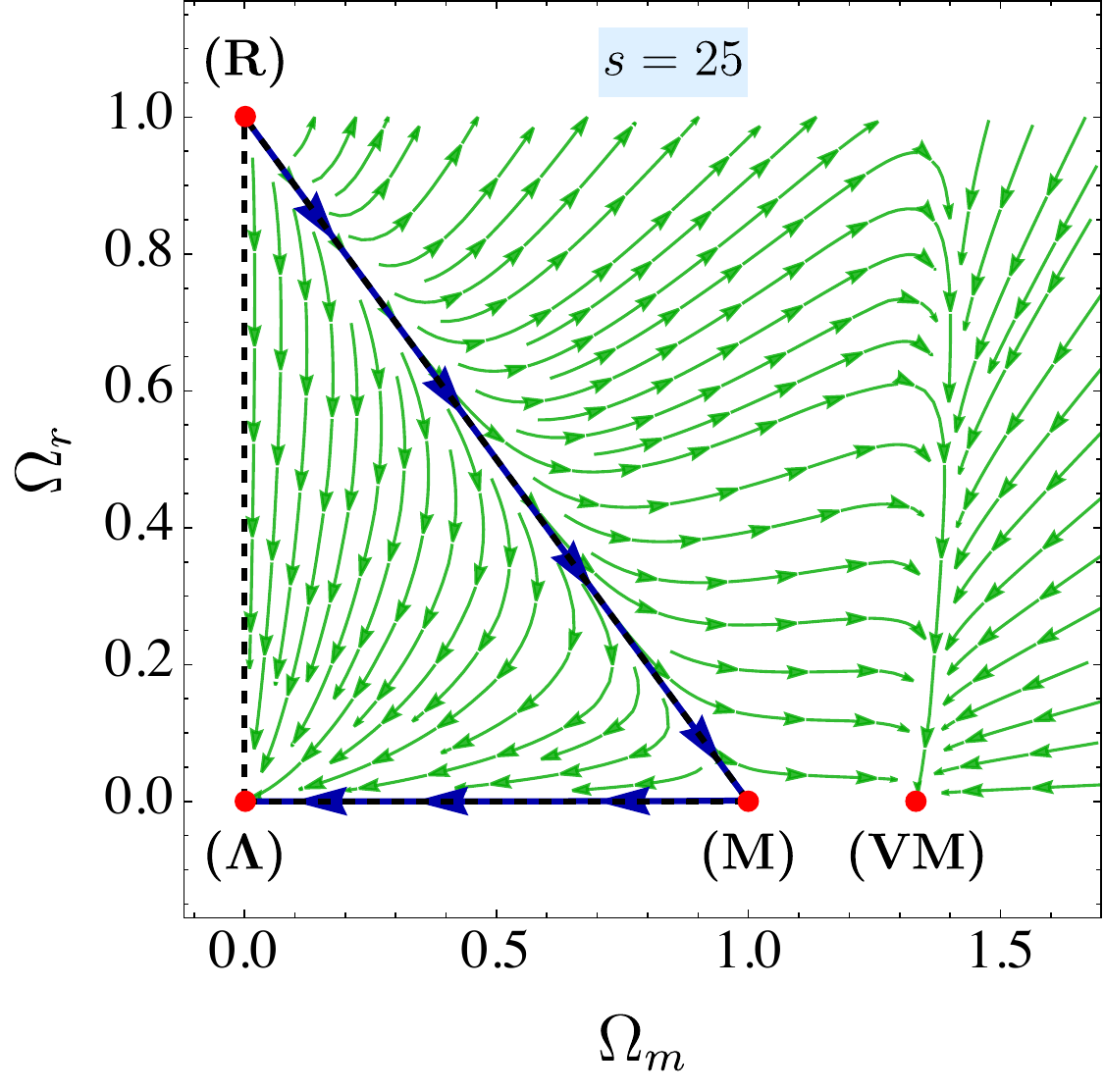}}
    \caption{Phase space portraits in the $(\Omega_m, \Omega_r)$ plane for the power-law viscous parameterization with $\hat{\xi}_0 = 10^{-3}$, together with the cosmological solution corresponding to the initial conditions of Eq.~\eqref{Eq: ICs} (solid blue line). \textbf{Left:} $s = 1.2$. The dynamical flow confines the cosmological trajectories within the invariant region bounded by the radiation, matter, and de Sitter fixed points. \textbf{Right:} $s = 25$. The viscous matter critical point (VM) appears as an attractor situated at $\Omega_m \approx 1.33$. Since it is located outside the physically allowed region, trajectories starting within the physical phase space cannot approach it and are instead guided towards the standard de Sitter ($\Lambda$) node.}
    \label{Fig: Phase Space s > 1}
\end{figure*}

To check the stability results obtained in the preceding sections, we solve the autonomous dynamical system numerically. Typically, initial conditions are anchored at the present epoch ($N=0$), where the density parameters are typically set to:
\begin{equation}
    \Omega_{r, 0} \approx 10^{-4}, \quad \Omega_{m,0} \approx 0.3, \quad \Omega_{\Lambda, 0} \approx 0.7.
\end{equation} 
However, a structurally complete cosmological trajectory should begin from the primordial radiation-dominated node, pass through an intermediate matter-dominated phase which should last enough to account for the structure formation, and then approach to the late-time accelerated attractor.

As shown in section~\ref{sa}, the stability analysis yielded the condition $s > 1$ to prevent divergent behaviour in the neighborhood of the radiation fixed point. Furthermore, to preserve a standard matter-dominated epoch consistent with structure formation, the dimensionless viscosity parameter must be small ($\hat{\xi}_0 \ll 1$). These restrictions indicate that bulk viscous effects must remain negligible throughout these epochs, which means that the early-universe background dynamics are close to those of the standard $\Lambda$CDM model. This behaviour enables the forward evolution to be initialized deep within the radiation era by reversing the present-day observational densities using the standard background Friedmann evolution:
\begin{equation}
    \left[\frac{H(N)}{H_0}\right]^2 = \Omega_{r, 0} e^{-4N} + \Omega_{m, 0} e^{-3N} + \Omega_{\Lambda, 0}.
\end{equation}
Setting the initial integration point at $N_i = -20$ (corresponding to a redshift $z \approx 4.85 \times 10^8$), the density parameters are:
\begin{align}
\label{Eq: ICs}
    \Omega_{r, i} &\approx 1, \nonumber \\
    \Omega_{m, i} &\approx 6.87 \times 10^{-6}, \nonumber \\ \Omega_{\Lambda, i} &\approx 1.4 \times 10^{-31}.
\end{align}

As the initial value of the density parameter related to the cosmological constant is extremely small at early times, standard double-precision arithmetic is susceptible to underflow or truncation errors when carrying out back propagation. To achieve numerical stability and to properly represent the cosmological transitions from the beginning up to the late-time attractor, the numerical integration is performed utilizing arbitrary-precision algorithms (keeping 40 significant digits), evolving the system continuously from the deep radiation era ($N = -38$) to the distant future ($N = 30$).

\subsubsection{\textbf{Case:} $s > 1$}

\begin{figure*}[t]
    \centering
    {\includegraphics[width=0.48\linewidth]{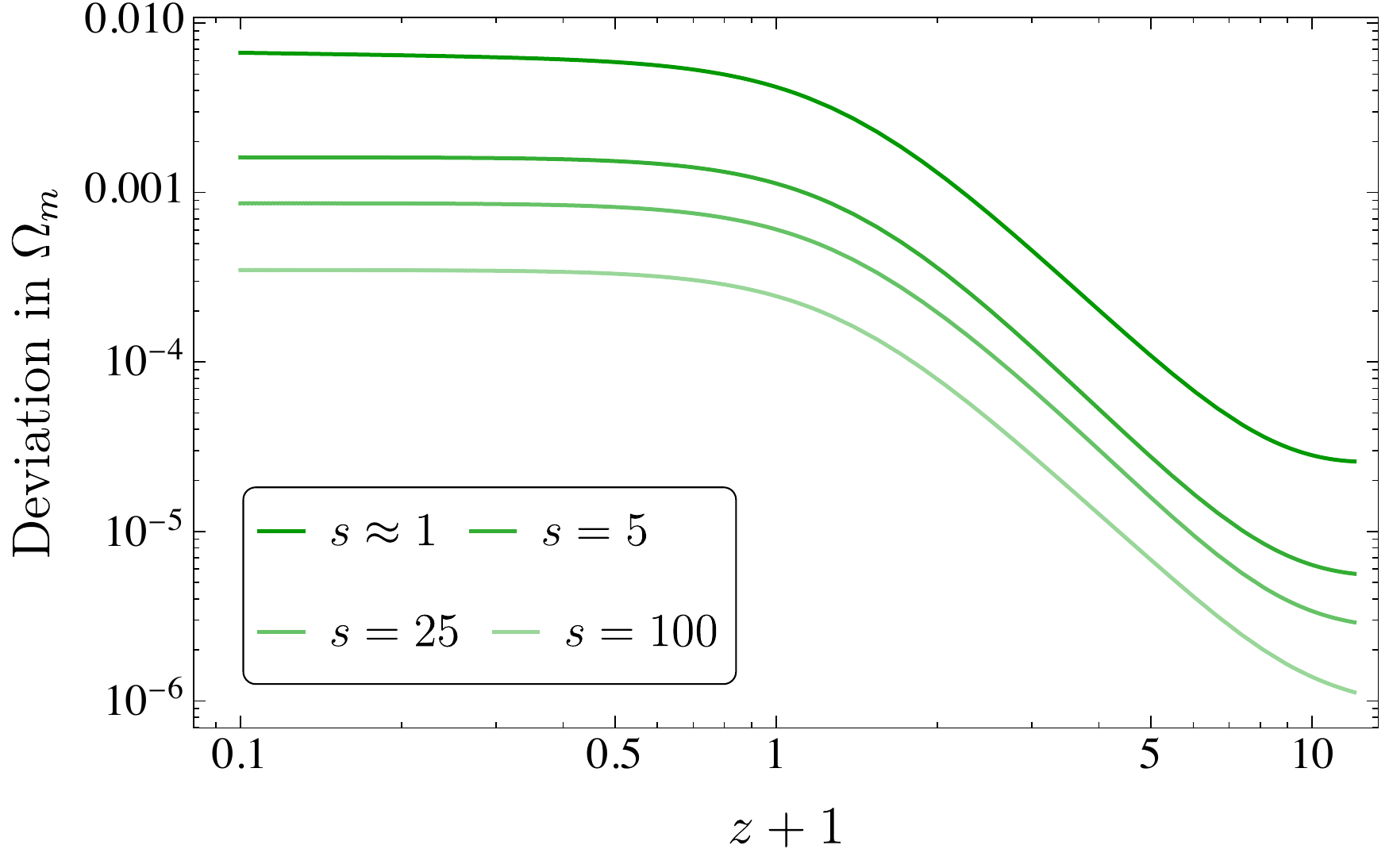} \hfill
    \includegraphics[width=0.48\linewidth]{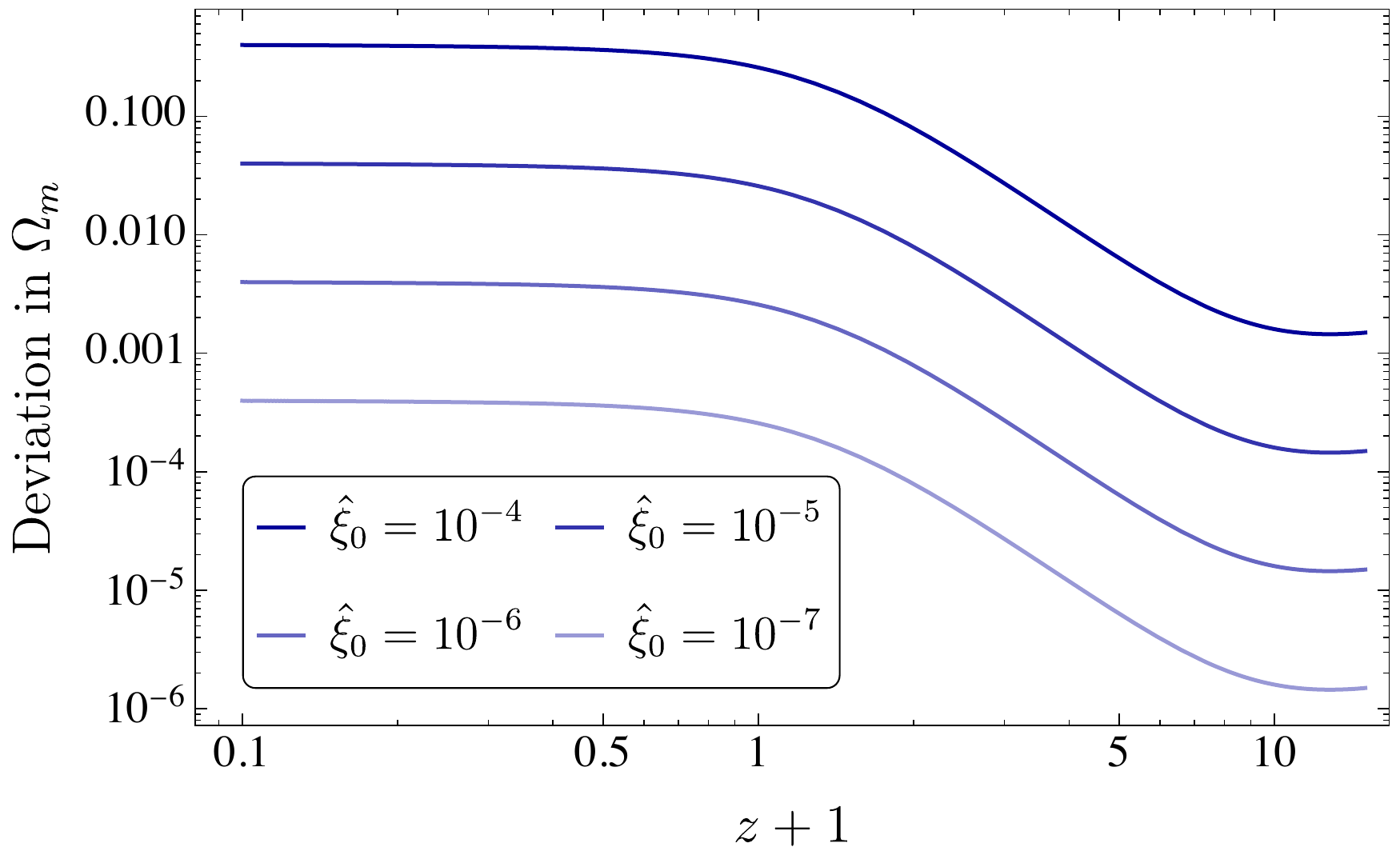}}
    \caption{Redshift evolution of the percentage difference in  $\Omega_m$ between the viscous model and the standard $\Lambda$CDM baseline calculated as $|1 - \frac{\Omega_{\text{m,viscous}}}{\Omega_{m,\Lambda\text{CDM}}}| \times 100$. \textbf{Left:} Variation of the power $s$ (fixing $\hat{\xi}_0 = 10^{-4}$). \textbf{Right:} Variation of the dimensionless viscosity parameter $\hat{\xi}_0$ (fixing $s = 1.2$). Notably, the deviation stays strictly under $1\%$ from $z = 10$ to the present day, which shows the strong background degeneracy between the two models.}
    \label{Fig: Abundances s > 1}
\end{figure*}

To verify the analytical constraints identified in the local stability analysis, we consider the realization with $s = 1.2$ and a small dimensionless viscosity parameter $\hat{\xi}_0 = 10^{-3}$. At the viscous matter point (VM), the equilibrium condition gives $\Omega_m = \hat{\xi}_0^{1/(1-s)} = 10^{15}$. Consequently, this attractor is moved out to values that lie well beyond the physically feasible phase space. The left panel of Fig.~\ref{Fig: Phase Space s > 1} displays this dynamical flow, showing that the physical cosmological trajectory stays within the limits of the physical phase space.

In the early radiation-dominated period ($q \approx 1$), the evolution equation for the matter density parameter simplifies to:
\begin{equation}
\label{Eq: Om_s_1_N}
    \frac{\dd \Omega_m}{\dd N} \approx \Omega_m + 3 \hat{\xi}_0 \Omega_m^s.
\end{equation}
Since $s > 1$ and $\Omega_m \ll 1$ in the early universe, the power of the right-hand-side of Eq. \eqref{Eq: Om_s_1_N} is subdominant ($\Omega_m^s \ll \Omega_m$), and the equation can be approximated to $\dd \Omega_m/\dd N \approx \Omega_m$. The system thus naturally reproduces the standard $\Lambda$CDM background behaviour, as depicted in Fig.~\ref{Fig: Abundances s > 1}. Both panels display the percentage difference in the evolution of $\Omega_m$ between the viscous framework and the baseline model, illustrating that the relative discrepancy stays strictly below $1\%$.

On the other hand, to illustrate the existence of the (VM) configuration as a stable node, we consider $s = 25$ and retain $\hat{\xi}_0 = 10^{-3}$. This yields an equilibrium coordinate of $\Omega_m \approx 1.33$, so that the attractor lies outside the physical boundaries but still within the visible part of the phase space portrait, as can be seen in the right panel of Fig.~\ref{Fig: Phase Space s > 1}. Even though the vector flow identifies (VM) as a stable node, the cosmological trajectory cannot cross the invariant boundaries of the physical subspace ($\Omega_r + \Omega_m \leq 1$). Instead, the flow is driven toward the standard non-viscous de Sitter attractor ($\Lambda$).

Conversely, during the intermediate matter-dominated era where $q \approx 1/2$, the evolution equation simplifies to:
\begin{equation}
    \frac{\dd \Omega_m}{\dd N} \approx 3 \hat{\xi}_0 \Omega_m^s.
\end{equation}
Since $\Omega_m \approx 1$ and $\hat{\xi}_0 \ll 1$, the matter density remains approximately constant and thus keeps the expected properties of a pressureless dust universe. However, the viscosity influences the effective equation of state, $w_\text{eff} \approx - \hat{\xi}_0$, at this point. As a result, an increased $\hat{\xi}_0$ causes premature cosmic acceleration, markedly shortening the matter-dominated epoch and impacting the formation of large-scale structure.

Within the standard cosmological model, the matter-dominated period roughly extends from the matter-radiation equality epoch ($z_\text{eq} \approx 3200$) to the point at which dark energy begins to dominate ($z_\text{acc} \approx 0.7$), which amounts to $N_m \approx 8$ $e$-folds. Numerical integration shows that if $s = 1.2$ and the viscosity parameter is increased to $\hat{\xi}_0 = 0.01$, the matter era is reduced to $N_m \approx 4$ $e$-folds. As an estimation of the possible effects of such a short matter-dominated epoch, we recall that the matter density perturbations scale as $\delta_m(a) \propto a$ in the linear perturbation regime~\cite{Bernardeau:2001qr,Mo:2010}. It follows, that the ratio $R_D$ of the growth factor between the ends of these eras can be estimated as:
\begin{equation}
    R_D \equiv \frac{\delta_m(a_\text{acc})}{\delta_m(a_\text{eq})} = e^{N_m}.
\end{equation}
In the case of $\Lambda$CDM ($N_m \approx 8$), the growth ratio is $R_D \approx 3 \times 10^{3}$, whereas in the viscous case with $N_m \approx 4$, it is approximately $R_D \approx 55$. At the time of radiation-matter equality, a primordial perturbation of amplitude $\delta_{m}(a_\text{eq}) \sim 10^{-5}$, as found from CMB observations~\cite{Planck:2018vyg}, grows to $\delta_m \sim 10^{-2}$ in the standard scenario, thus bringing them into the range required for non-linear gravitational collapse~\cite{Mo:2010}. In contrast, the shortened matter era in the viscous model leads to $\delta_m \sim 5 \times 10^{-4}$, which is not enough to explain the  large-scale structure of the universe.

The fact that the condition $s > 1$ must be guaranteed to ensure stable past and future asymptotic states, together with the observational constraints from structure formation, means that the bulk viscosity parameter is confined to be very small. As a result, the background cosmological evolution of the model becomes effectively indistinguishable from the standard $\Lambda$CDM model.

\subsubsection{\textbf{Case:} $s < 1$}

\begin{figure*}[t]
    \centering
    {\includegraphics[width=0.35\linewidth]{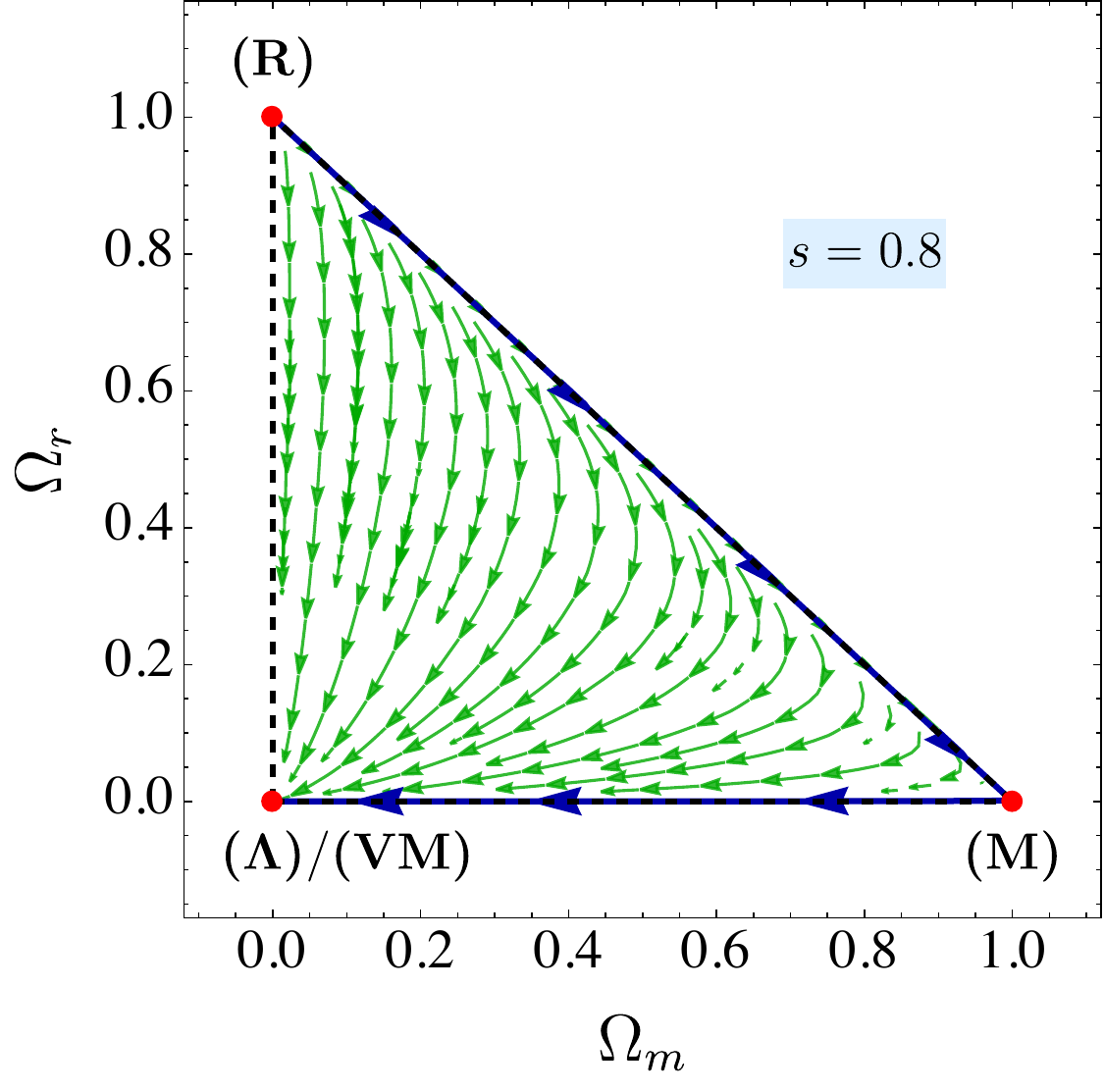} \hspace{2cm}
    \includegraphics[width=0.365\linewidth]{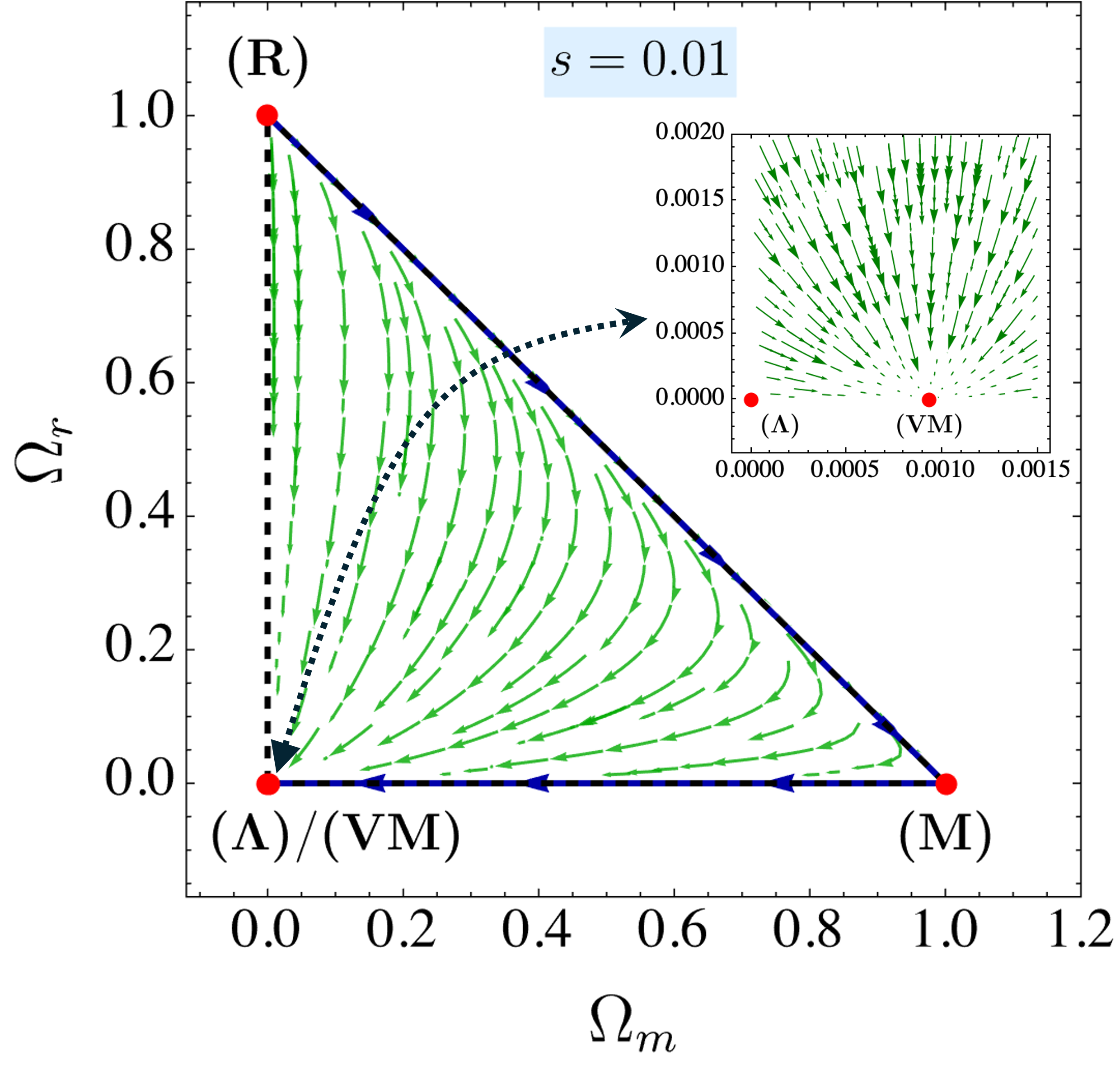}}
    \caption{Phase space dynamics for small viscous parameters ($s < 1$). \textbf{Left panel:} For $s = 0.8$, the attractor (VM) and the saddle ($\Lambda$) are closely situated in the phase space, giving an approximation to standard late-time cosmology yet showing past-time integration constraints. \textbf{Right panel:} In the alternative case $s = 0.01$, the topological separation between the ($\Lambda$) saddle and the true (VM) attractor becomes noticeable (as indicated in the zoomed inset vector field). Nevertheless, cosmological trajectories end in the past region because of the transition of the dark matter density into negative regimes where fractional powers result in complex values.}
    \label{Fig: Phase Space s < 1}
\end{figure*}

Now, we address the $s<1$ case. For this region of the parameter space, the stability is opposite to the previous case. The viscous matter point (VM) becomes a stable late-time attractor, while the de Sitter Point ($\Lambda$) is a saddle. Although in this region the viscous pressure can sustain a late-time accelerated expansion, there is a challenge in establishing an early radiation-dominated epoch, manifesting as a past-time boundary beyond which backward integration cannot be extended.

Assuming $s < 1$ and $\hat{\xi}_0 \ll 1$ for a viable matter-dominated era, the deceleration parameter in the early radiation-domination is $q \approx 1$. Consequently, the main terms in the evolution equation for the matter density parameter become those of a Bernoulli differential equation~\cite{Boyce2017}:
\begin{equation}
    \frac{\dd \Omega_m}{\dd N} \approx \Omega_m + 3\hat{\xi}_0\Omega_m^s,
\end{equation}
which can be analytically integrated. By integrating backward from a given initial condition $\Omega_{m, i}$ at an intermediate $e$-fold $N_i$, we get:
\begin{equation}
    \Omega_m^{1-s} = (\Omega_{m, i}^{1 - s} + 3 \hat{\xi}_0)e^{(1-s)(N - N_i)} - 3\hat{\xi}_0.
\end{equation}
The analytical solution shows a structure limitation in the case where $s < 1$. If the integration is carried out in the backward direction ($N < N_i$), the exponential factor decreases monotonically. In consequence, the right-hand side reaches zero at some finite value, below which $\Omega_m$ becomes negative. Since the dimensionless viscous pressure depends on a power of the dark matter density ($\Omega_\Pi = - \hat{\xi}_0\Omega_m^s$), negative values of $\Omega_m$ correspond to points in the complex plane and, therefore, they give rise to unphysical values, so a lower bound is imposed in the integration. We define this precise integration threshold as $N_\text{crash}$, given by:
\begin{equation}
    N_\text{crash} \equiv N_i - \frac{1}{1-s}\ln\left(1 + \frac{\Omega_{m,i}^{1-s}}{3 \hat{\xi}_0} \right).
\end{equation}
The behaviour of this boundary shows the parametric trade-offs present in this regime. In the limit $s \to 1$, $N_{\text{crash}} \to -\infty$, eliminating the numerical termination; however, this pushes the (VM) equilibrium coordinate ($\Omega_m = \hat{\xi}_0^{1/(1-s)}$) toward zero, and the dissipative effect is suppressed. On the other hand, in the limit $s \to 0$, the contribution from the late-time viscosity is increased while $N_{\text{crash}} \to N_i$ simultaneously, reducing the backward integration range and restricting the accessible early-time state space.

To illustrate this behaviour, we take a baseline value of $\hat{\xi}_0 = 10^{-3}$. For $s = 0.8$, the theoretical threshold occurs at $N_{\text{crash}} \approx -37.3$. As shown in Table~\ref{tab:crash}, the numerical values across this boundary reveal the transition of the variables into the complex domain over a narrow range of $e$-folds.

\begin{figure}[t]
    \centering
    \includegraphics[width=0.8\linewidth]{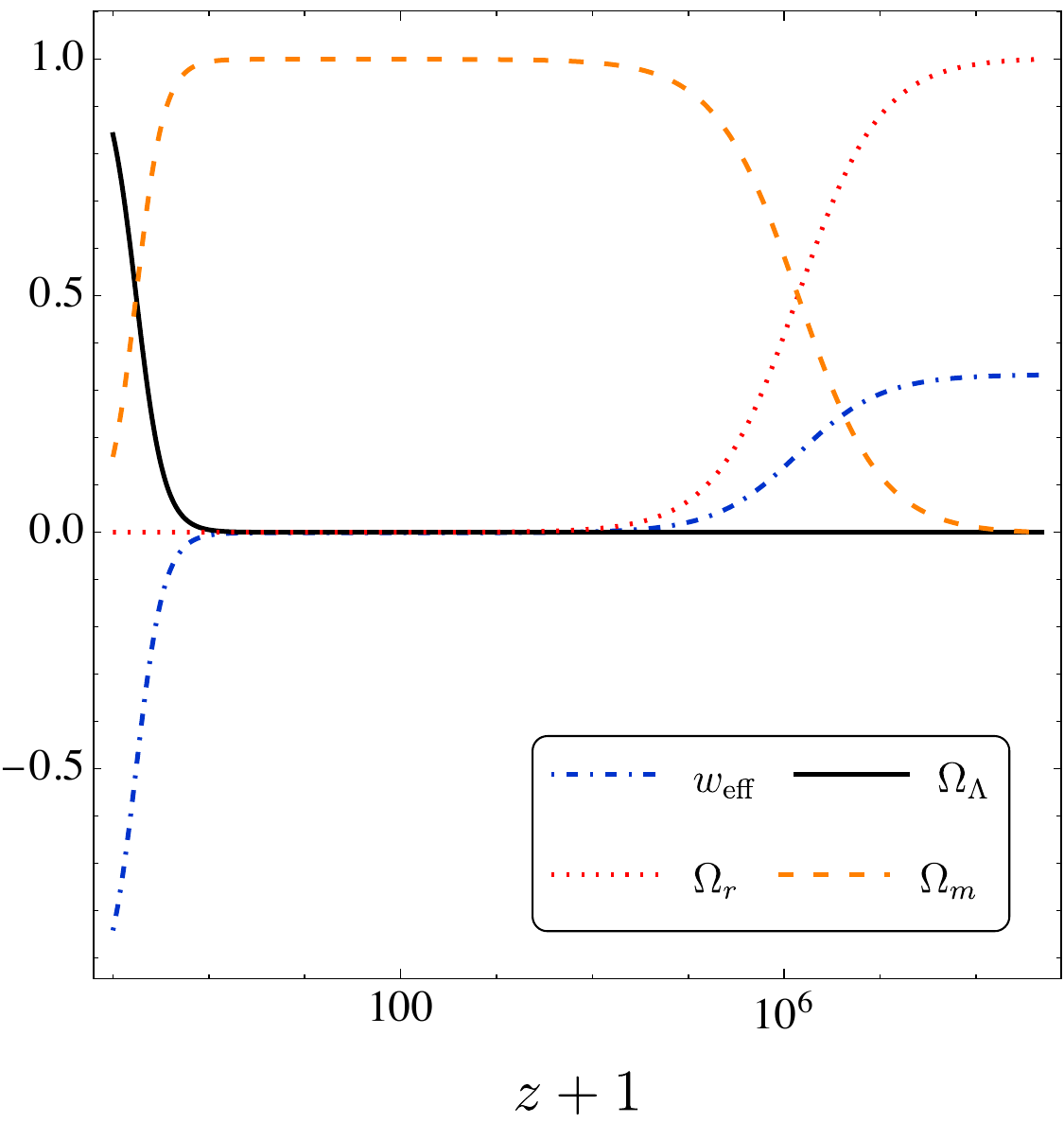}
    \caption{Cosmological evolution of the density parameters and the effective equation of state assuming $s = 0.01$ and $\hat{\xi}_0 = 10^{-3}$. Although the (VM) point acts as the late-time attractor, the exponent modifies the intermediate cosmic evolution by prolonging the matter-dominated epoch. Moreover, backward numerical integration stops at $N_{\text{crash}} \approx -20$ (corresponding to $z \approx 5 \times 10^8$), showing that a deep radiation era cannot be established within this regime.}
    \label{Fig: Abundance s < 1}
\end{figure}

Under these conditions, the (VM) attractor is dynamically close to ($\Lambda$), as depicted in the left panel of Fig.~\ref{Fig: Phase Space s < 1}. To distinguish these critical points, we choose $s = 0.01$; this shifts the integration barrier closer to the present epoch, yielding $N_{\text{crash}} \approx -20$. Numerically, $\Omega_m$ vanishes at $N \approx -20.002$ and, by $N = -21$, the viscous pressure acquires a non-vanishing imaginary part (see Table~\ref{tab:crash}). While the true (VM) attractor becomes topologically distinguishable (right panel of Fig.~\ref{Fig: Phase Space s < 1}), this region of the parameter space induces an extended matter-dominated era (Fig.~\ref{Fig: Abundance s < 1}) and limits the viability of a long enough radiation-dominated epoch as calculated in Ref.~\cite{Alvarez:2019ues}.

\begin{table}[ht!]
\centering
\caption{Evolution of the matter density and viscous pressure across the past-time integration boundary for $\hat{\xi}_0 = 10^{-3}$.}
\label{tab:crash}
\begin{tabular}{c | c | c}
\hline\hline
Model Case & $N$ & $\Omega_\Pi$ \\
\hline
$s = 0.8$ & $-37.29$ & $2.34 \times 10^{-21}$ \\ 
\hline
$s = 0.8$ & $-37.30$ & $-1.26 \times 10^{-21} - 1.08 \times 10^{-21} \, i$ \\
\hline
$s = 0.01$ & $-20$ & $0.0027$ \\ 
\hline
$s = 0.01$ & $-21$ & $0.0077 + 2.75 \times 10^{-6} \, i$ \\
\hline\hline
\end{tabular}
\end{table}

\subsection{Crucial Constraints from Linear Perturbations: The Sub-Horizon Limit}
\label{Sec: Perturbations}

While we have demonstrated that the background cosmology for $s \geq 1$ and $\hat{\xi}_0 \ll 1$ closely tracks the standard $\Lambda$CDM evolution, a more rigorous assessment requires verifying the stability and growth of linear perturbations. In models where the background dynamics mimics the standard model, the primary gauge of physical viability shifts to the growth of cosmic structures~\cite{Li:2009mf, Velten:2014xca, Blas:2015tla, Barbosa:2017ojt, Arjona:2018jhh, Arjona:2019rfn, Cardona:2022lcz}.

Let us examine the linear evolution of viscous dark matter perturbations. The full perturbative equations for scales deep inside the Hubble horizon are given in Appendix~\ref{App: Linear Theory}. In the sub-horizon limit the governing equation for the matter density contrast $\delta_m \equiv \delta\rho_m / \bar{\rho}_m$ becomes:
\begin{equation}
\label{Eq: General delta eq}
    \delta_m'' + \mathcal{H} \delta_m' + k^2 \frac{\delta\Pi}{\bar\rho_m} \approx 4\pi G a^2 \bar\rho_m \delta_m,
\end{equation}
where primes denote derivatives with respect to conformal time $\tau$, $\mathcal{H} \equiv a'/a$ is the conformal Hubble parameter which is related to the Hubble rate by $\mathcal{H} = a H$, $\delta\Pi$ stands for the first-order perturbation of the bulk viscous pressure, and an overbar denotes that the quantity is evaluated at the background level. In the small-scale regime ($k \gg H$), the first-order viscous pressure perturbation satisfies the relation:
\begin{equation}
\label{Eq: Perturbed Viscosity}
    \delta\Pi \approx \xi \, \delta'_m.
\end{equation}
Substituting this expression directly into Eq.~\eqref{Eq: General delta eq} results in an enhanced friction term:
\begin{equation}
    \delta_m'' + \mathcal{H} \left(1 + R_\text{v} \right) \delta_m' - 4\pi G a^2 \bar\rho_m \delta_m \approx 0, 
\end{equation}
where the dimensionless scale-dependent parameter $R_\text{v}$ is defined as:
\begin{equation}
    R_\text{v} \equiv \left(\frac{k^2}{\mathcal{H}^2}\right) \left(\frac{\mathcal{H}}{a} \right) \frac{\xi}{\bar\rho_m}.
\end{equation}
Consequently, the bulk viscosity gives rise to an effective dissipative drag, which restricts the growth of structures on sub-horizon scales. To evaluate the magnitude of this effect, we apply the power-law parameterization to the Eckart regime which, in the matter-dominated epoch ($\Omega_m \approx 1$), yields:
\begin{equation}
    \frac{\xi}{\bar\rho_m} \approx \frac{\hat{\xi}_0}{3H} \ \implies \ R_\text{v} \approx \left( \frac{k}{H_0} \right)^2 \frac{\hat{\xi}_0}{3\Omega_{m,0}} a.
\end{equation}
For $a \ll 1$ and under the background constraint $\hat{\xi}_0 \ll 1$, the additional viscous friction remains highly subdominant ($R_\text{v} \ll 1$). However, on small scales this suppression could be large due to the $k^2$ dependence; this is a remnant of the gradient instabilities in Eckart theory~\cite{Hiscock:1985zz}.

We can track the deviations from the standard paradigm. We rewrite the perturbation equation in terms of the scale factor $a$ as:
\begin{equation}
    \frac{\dd^2 \delta_m}{\dd a^2} + \left(\frac{3}{2a} + \frac{R_\text{v}}{a}\right) \frac{\dd \delta_m}{\dd a} - \frac{3}{2a^2} \delta_m = 0.
\end{equation}
In the standard $\Lambda$CDM model ($R_\text{v} \to 0$), the growing mode is given by $\delta_\Lambda(a) \propto a$. We treat the viscous solution as a first-order correction around this baseline:
\begin{equation}
    \delta_m (a, k) = a + \epsilon(a, k), \quad \epsilon(a, k) \ll a.
\end{equation}
Substituting this ansatz into the differential equation yields the evolution equation for the deviation $\epsilon(a)$:
\begin{equation}
    \frac{\dd^2 \epsilon}{\dd a^2} + \frac{3}{2a}\frac{\dd \epsilon}{\dd a} - \frac{3\epsilon}{2a^2} = - \tilde{R}_\text{v}(k),
\end{equation}
where 
\begin{equation}
    \tilde{R}_\text{v}(k) \equiv \left(k/H_0\right)^2 \left(\hat{\xi}_0 / 3\Omega_{m,0}\right).
\end{equation}
This inhomogeneous equation can be solved analytically, yielding the growing mode solution:
\begin{equation}
    \epsilon(a, k) = - \frac{2}{7} \tilde{R}_\text{v}(k) a^3,
\end{equation}
which satisfies the condition $\epsilon \ll a$ whenever $a \leq 1$. If the matter power spectrum at the present day ($a = 1$) is $P(k) \propto \delta_m^2(k)$, the fractional deviation from the standard $\Lambda$CDM spectrum $P_\Lambda(k)$ reads:
\begin{equation}
    \frac{\Delta P}{P_\Lambda} \equiv \frac{P(k) - P_\Lambda(k)}{P_\Lambda(k)} \approx - \frac{2}{9} \bar{R}_\text{v}(k).
\end{equation}
Using a typical linear scale $k = 300 H_0 \approx 0.1~h\,\text{Mpc}^{-1}$, and adopting the conservative value $\hat{\xi}_0 = 10^{-6}$ according to Ref.~\cite{Anand:2017wsj}, alongside $\Omega_{m, 0} = 0.3$, the power spectrum is suppressed by about $\Delta P/P_\Lambda \approx -2.5\%$. At slightly smaller scales where $k = 600 H_0 \approx 0.2~h\,\text{Mpc}^{-1}$, however, the suppression becomes considerable, reaching $\approx -23\%$, and our perturbative treatment around the $\Lambda$CDM baseline is no longer reliable. The fact that this effect depends on scale in such a strict $k^2$ manner is particularly noteworthy since the model has almost no impact on the large-scale CMB observations while actively reducing the growth of structure on smaller scales. Although this is a promising feature, standard linear theory fails at these high wavenumbers, so a definite verification of this mechanism will have to wait until the analysis is extended into the mildly non-linear regime.

In a setting where $s > 1$, the ratio $\xi/\bar{\rho}_m$ decreases more rapidly since $\Omega_m^{s - 1} \propto a^{-3(s-1)}$ during matter domination. This causes the viscous drag manifests at early times. This analytical finding demonstrates that within the background-mimicking regime ($\hat{\xi}_0 \ll 1$), the accumulated bulk viscosity could offer a natural mechanism to suppress structure growth. The region of parameter space which can appear as $\Lambda$CDM at the background level still shows a quite noticeable phenomenological effect at perturbative scales, even though it is unlikely to alleviate the $H_0$ tension. 

\section{Discussion and Conclusions}
\label{Sec: Conclusions}

In this work, we evaluated the cosmological feasibility of bulk viscous dark matter as a mechanism to reproduce a consistent cosmological history in the background and modulate the growth of cosmic structures. 

By adopting the rather general restriction $\Omega_\Pi = \Omega_\Pi(\Omega_m)$, which encompasses a broad class of bulk viscous models considered in the literature, together with the negative bulk viscous pressure obtained in the regime of validity of the Eckart theory, we have demonstrated that the phase-space structure itself imposes robust constraints:
\begin{itemize}
    \item \textbf{Early-Universe behaviour:} We have found that the only mathematical possibility for a viable radiation-dominated source node is the complete absence of early-time viscosity. To properly initiate a standard cosmic chronology, the Universe must begin its evolution as a standard $\Lambda$CDM framework; bulk viscosity is fundamentally restricted to a late-time phenomenon. A similar result was claimed in Ref.~\cite{Sasidharan:2015ihq} for particular parameterizations of the bulk viscosity.
    \item \textbf{The Matter-Dominated Epoch:} In order to have a matter-dominated epoch that lasts for a sufficiently long time to agree with large-scale structure observations, the viscous pressure must be extremely small. In consequence, the background cosmology must closely track the standard $\Lambda$CDM evolution during both the radiation and matter eras.
    \item \textbf{Late-Time Attractors:} Although a dedicated Viscous Matter (VM) fixed point exists and is capable of acting as an attractor that drives late-time accelerated expansion, a rigorous assessment reveals that this point lies outside the physical phase space of the theory. Instead, the only stable, physically admissible attractor available to govern the asymptotic future of the cosmos is the standard cosmological constant ($\Lambda$)-dominated fixed point.
\end{itemize}

To establish these constraints numerically, we assumed a widely employed power-law parameterization where the bulk viscosity scales with the matter density parameter as $\Omega_\Pi = -\hat{\xi}_0 \Omega_m^s$. The resulting stability analysis and numerical integrations showed a clear parametric dichotomy. If $s \geq 1$ and the dimensionless viscosity parameter $\hat{\xi}_0 \ll 1$, the past-time limits remain mathematically well-behaved, but the background dynamics becomes practically indistinguishable from the standard $\Lambda$CDM. On the other hand, in the case where $s < 1$, the (VM) fixed point acts as the stable late-time attractor which is virtually identical to the ($\Lambda$) baseline. Yet, this regime leads to an integration boundary in the past, characterized by a finite scale factor threshold $N_\text{crash}$. Beyond this point toward the past, the dark matter density takes on negative values, causing $\Omega_\Pi$ to become a complex number, and effectively rules out the possibility of having a sufficiently long primordial radiation era. These structural constraints therefore offer useful background information for the previous numerical studies mentioned in the literature. As we suggest in Appendix~\ref{App: Truncated IsrSte}, these results may extend to a truncated Israel--Stewart theory. Nonetheless, a rigorous study of these features will be the subject of a future investigation.

Since the background evolution in the case of viable $s \geq 1$ must essentially reproduce the standard $\Lambda$CDM expansion history, it is unlikely that bulk viscous dark matter models can provide a real solution to the $H_0$ tension. Nevertheless, a linear perturbative treatment under the sub-horizon and quasi-static approximations reveals that the viscosity causes an effective scale-dependent dissipative drag. This effect enhances the Hubble friction, thus providing a possible way of suppressing the growth of structure. However, since the viscous drag introduces a strong scale dependence, a standard linear approximation provides an incomplete description. A more thorough, scale-by-scale linear analysis is therefore strictly necessary to fully characterize how this model affects the matter power spectrum, a task which we will leave for future study.

The mathematical boundaries identified in this study—particularly the past-time integration limits driven by viscous models––suggest that standard, non-viscous interacting dark sector models~\cite{Boehmer:2008av, Caldera-Cabral:2008yyo, Wang:2016lxa} featuring an energy exchange rate $Q \propto \rho^s$ might suffer from analogous structural defects. We conjecture that such interacting models could face similar challenges in simultaneously preserving a stable primordial radiation era and an adequately prolonged matter-dominated epoch. A formal dynamical-systems analysis of these interacting scenarios to verify these potential features will be the subject of future investigation. 

Extending the scope of our research to the level of full Israel--Stewart theories will also be the subject of forthcoming work.

\section*{Acknowledgements}

J. Bayron Orjuela-Quintana has been funded by Vicerrector\'ia de Investigaciones
- Universidad del Valle grant number 71394. Yeinzon Rodríguez has been funded by Universidad Antonio Nariño under grant numbers VCTI 2024211, VCTI 2025205, VCTI 2232026, and VCTI 2252026. We acknowledge Gabriel Gómez for clarifying discussions and interesting feedback on our work.

\textbf{Declaration of AI Assistance:} The authors declare the AI assistance to improve the language and readability of the paper. The authors have reviewed and edited the content and take full responsibility for the content of the manuscript.

\appendix

\section{Negative Bulk Viscous Pressure in the Truncated Israel--Stewart Theory}
\label{App: Truncated IsrSte}

The central arguments that backup our findings are the assumption $\Omega_\Pi= \Omega_\Pi(\Omega_m)$ and the negativity of the bulk viscous pressure. We show that, if we retain that general form for $\Omega_\Pi$, the bulk viscous pressure tends to negative values in the truncated Israel--Stewart (IS) theory.

In the Eckart formalism, the bulk viscous pressure is given by $\Pi = -3H\xi$. Since $\xi>0$, the value of $\Omega_\Pi$ is always negative. In the case of the truncated IS formalism, $\Pi$ evolves according to:
\begin{equation}
    \Pi + \tau_\mathrm{r} \, \dot{\Pi} = - 3 H \xi,
    \label{eq: ApTIS}
\end{equation}
where $\tau_\mathrm{r} \geq 0$ is the relaxation time. 

To constrain the sign of $\Omega_\Pi$, we replace the definition of $\Omega_\Pi$ [Eq.~\eqref{Eq: Omegapi}] in Eq.~\eqref{eq: ApTIS}. Then, we get an independent evolution equation for $\Omega_\Pi$:
\begin{equation}
    \frac{\dd \Omega_\Pi}{\dd N} = \Omega_\Pi \left[ 2(q+1) - \frac{1}{H\tau_\mathrm{r}}\right] - \frac{8\pi G_\mathrm{N}}{H \tau_\mathrm{r}} \left(\frac{\xi}{H} \right).
\end{equation}
In a fixed point, $\dd \Omega_\Pi / \dd N = 0$, so the bulk viscous pressure can be written as:
\begin{equation}
    \frac{\Pi}{3H\xi} = \frac{1}{2(q + 1) H\tau_\mathrm{r} -1}.
\end{equation}
In the regime of validity of the truncated IS approximation ($H \tau_r \ll 1$)~\cite{Maartens:1996vi, Zimdahl:1996ka}, the bulk viscous pressure tends to negative values. This suggests that the findings in this work could be extended to the truncated IS formalism. This case will be left for forthcoming research.

\section{Cosmological Perturbation Theory for Bulk Viscous Dark Matter}
\label{App: Linear Theory}

To asses the suppression of structure growth caused by the dissipation of dark matter, we introduce scalar perturbations to the homogeneous and isotropic background. The perturbed space-time metric is given in the conformal Newtonian gauge as:
\begin{equation}
    \dd s^2 = a^2(\tau)\left[ - (1 + 2 \Psi) \dd \tau^2 + (1 + 2\Phi)\dd x_i \dd x^i \right],
\end{equation}
where $\Psi$ and $\Phi$ are the gravitational potentials, and $\tau$ denotes the conformal time. Primes indicate derivatives with respect $\tau$, and $\mathcal{H} \equiv a'/a$ defines the conformal Hubble parameter.

For a cosmic fluid characterized by an energy density contrast $\delta \equiv \delta\rho/\bar\rho$ and a velocity divergence $\theta \equiv \partial_i v^i$, the conservation of the energy-momentum tensor ($\nabla_\mu T^{\mu\nu} = 0$) leads to the usual first-order fluid evolution equations in Fourier space:
\begin{align}
    \delta' = &- (1 + w) (\theta + 3\Phi') - 3 \mathcal{H}(c_s^2 - w) \delta, \\
    \theta' = &- \mathcal{H}(1 - 3w) \theta - \frac{w'}{1 + w} \theta + \frac{c_s^2}{1+w} k^2 \delta \nonumber \\
    &- k^2 \sigma + k^2 \Psi,
\end{align}
where $w \equiv \bar p/\bar\rho$ is the equation of state of the fluid, $c_s^2 \equiv \delta p/\delta \rho$ is its sound speed, $\sigma$ stands for its anisotropic stress, and $k$ is the wavenumber. 

Within the context of bulk viscous dark matter, the pressure is completely determined by the dissipative bulk viscous contribution, $p_m = \Pi$. Because a bulk viscous fluid is incapable of supporting directional shearing layer forces at the linear level, the anisotropic stress vanishes identically. Therefore, the viscous fluid parameters are given by:
\begin{equation}
    w_m = \frac{\bar{\Pi}}{\bar\rho_m}, \quad c_{s, m}^2 = \frac{\delta \Pi}{\delta \rho_m}, \quad \sigma_m = 0.
\end{equation}
We now consider the dynamics during the matter-dominated epoch (the fixed point (M) characterized in the main text). In this case, the background dark matter density dominates the cosmic budget ($\Omega_m \approx 1$) and the background bulk viscous pressure is small ($|\Omega_\Pi| \ll 1$), which means $w_m \approx 0$. When these conditions are substituted in the general conservation equations, the perturbed system simplifies to:
\begin{align}
    \delta_m' &= - \theta_m - 3\Phi' - 3 \mathcal{H}\frac{\delta \Pi}{\bar\rho_m}, \label{Eq: App delta_m prime} \\
    \theta_m' &= - \mathcal{H} \theta_m + k^2 \frac{\delta \Pi}{\bar\rho_m} + k^2 \Psi, \label{Eq: App theta_m prime}
\end{align}
where we have employed the identity $c_{s,m}^2 \delta_m = \delta\Pi/\bar{\rho}_m$.

To extract the physical behaviour on scales relevant to large-scale structure anomalies, we invoke two standard cosmological tracking simplifications~\cite{Orjuela-Quintana:2023zjm}:
\begin{itemize}
    \item \textbf{Sub-Horizon Approximation (SHA):} We are dealing with modes whose physical wavelengths are much smaller than the cosmic horizon, which means that the wavenumber satisfies $k \gg \mathcal{H}$. As a result, spatial gradient terms scaled by $k^2$ systematically dominate over terms proportional to $\mathcal{H}^2$ or background time derivatives.
    \item \textbf{Quasi-Static Approximation (QSA):} We assume that the time evolution of the metric potentials is solely due to the slow background expansion rather than fast dynamical oscillations; in this case, time derivatives of the potentials are negligible in comparison with their spatial gradients ($\Phi' \sim \Psi' \sim 0$ and $\mathcal{H}\Phi \sim \mathcal{H}\Psi \sim 0$) and compared to the fluid's evolutionary velocity ($\Phi' \ll \theta_m$).
\end{itemize}
Applying the SHA and QSA limits to Eq.~\eqref{Eq: App delta_m prime}, the higher-order expansion terms disappear and the continuity equation is simplified to $\delta_m' \approx - \theta_m$. Differentiating this equation with respect to $\tau$ once more and then substituting the result into the Euler equation~\eqref{Eq: App theta_m prime}, we obtain a closed second-order differential equation for the dark matter contrast:
\begin{equation}
    \delta_m'' + \mathcal{H}\delta_m' + k^2\frac{\delta\Pi}{\bar\rho_m} = - k^2 \Psi.
    \label{Eq: Second order delta baseline}
\end{equation}

To eliminate the gravitational potential $\Psi$, we inspect the scalar components of the perturbed Einstein field equations:
\begin{align}
    3\mathcal{H}(\mathcal{H}\Psi - \Phi') - k^2 \Phi &= -4\pi G a^2 \delta\rho_m, \\
    - \frac{1}{3} k^2 (\Phi + \Psi) &= 4\pi G a^2 (\bar\rho_m + \bar{\Pi}) \sigma_m.
\end{align}
Under the SHA ($k^2 \gg \mathcal{H}^2$) and QSA ($\Phi' \to 0$, $\mathcal{H}\Psi \to 0$), the first equation reduces to the usual relativistic Poisson equation. Moreover, since $\sigma_m = 0$, the second equation requires the two gravitational potentials to be equal and opposite. Thus, the gravitational field equations simplify to:
\begin{equation}
    k^2 \Phi = 4\pi G a^2 \bar\rho_m \delta_m, \quad \Psi = - \Phi.
\end{equation}
Substituting $-k^2\Psi = k^2\Phi = 4\pi G a^2 \bar\rho_m \delta_m$ back into Eq.~\eqref{Eq: Second order delta baseline} yields:
\begin{equation}
     \delta_m'' + \mathcal{H}\delta_m' + k^2\frac{\delta\Pi}{\bar\rho_m} = 4\pi G a^2 \bar\rho_m \delta_m.
     \label{Eq: App Intermediate Poisson}
\end{equation}

To close this system, a perturbative expression for the bulk viscous pressure $\delta\Pi$ is required. The full Israel--Stewart transport equation is given by:
\begin{equation}
    \Pi + \tau_{\text{r}} \dot{\Pi} = - \xi \Theta - \frac{\tau_{\text{r}}}{2} \Pi \left[ \Theta + \frac{\dot{\tau}_{\text{r}}}{\tau_{\text{r}}} - \frac{\dot{\xi}}{\xi} - \frac{\dot{T}}{T} \right].
\end{equation}
In the first order of cosmological perturbation theory ($\Theta = \bar{\Theta} + \delta\Theta$, $\Pi = \bar{\Pi} + \delta\Pi$, \ldots), and taking into account that the microscopic relaxation timescale is extremely small compared to the macroscopic age of the Universe ($H \tau_{\text{r}}  \ll 1$), the dynamical second-order terms become sub-dominant. The system then reaches a localized relationship:
\begin{equation}
    \delta\Pi = -\xi \, \delta\Theta.
\end{equation}
In conformal Newtonian gauge, the perturbation of the expansion scalar for the dark matter component is equal to $\delta\Theta = \theta_m / a$. Since the continuity relation is valid under SHA and QSA ($\theta_m = -\delta_m'$), the pressure perturbation becomes directly proportional to the rate of growth of the matter overdensity:
\begin{equation}
    \delta\Pi = \frac{\xi}{a} \delta_m'.
\end{equation}
This shows that, in the Newtonian-limit, the first-order causal perturbations naturally give rise to a localized dissipative structure similar to that of Eckart theory. When we substitute this closure relation into Eq.~\eqref{Eq: App Intermediate Poisson}, we obtain the final evolution equation for the linear matter contrast:
\begin{equation}
    \delta_m'' + \mathcal{H}(1 + R_\text{v}) \delta_m' - 4\pi G a^2 \bar\rho_m \delta_m = 0,
\end{equation}
where we have defined the dimensionless viscous damping ratio as:
\begin{equation}
    R_\text{v} \equiv \frac{k^2 \xi}{a \mathcal{H} \bar\rho_m} = \frac{k^2 \xi}{\mathcal{H}^2 \bar\rho_m} \left(\frac{\mathcal{H}}{a}\right).
\end{equation}
The quantity $R_\text{v}$ serves as a scale-dependent correction to the usual Hubble drag. Because $R_\text{v} \propto k^2$, dissipation results into the suppression of the formation of structure on small scales. This directly affects the late-time matter power spectrum.

To compute this deviation explicitly, we choose to work within the verified power-law parameterization evaluated at sub-horizon scales:
\begin{equation}
    \frac{\xi}{\bar\rho_m} = - \frac{\bar{\Pi}}{3H\bar\rho_m} = \frac{\hat\xi_0}{3H} \Omega_m^{s-1}. 
\end{equation}
During the peak of the matter-dominated era, the background approximation $\Omega_m \approx 1$ holds. However, across the full cosmic trajectory, the background matter density dilutes as $\Omega_m \propto a^{-3}$. Consequently, for the physically viable regime characterized by $s > 1$, the viscous drag modification $R_\text{v}$ is heavily damped as the Universe expands, ensuring that the model remains safely anchored to a well-behaved $\Lambda$CDM baseline at late times while leaving an active imprint on mildly non-linear scales.

\bibliography{bibliography}

\end{document}